\documentclass[preprint,a4paper,superscriptaddress]{revtex4-2}
\setcitestyle{super}

\usepackage{amsmath,amsfonts,amssymb}
\usepackage{geometry}
\usepackage{graphicx}
\usepackage{color}
\usepackage{xcolor}
\usepackage{empheq}
\usepackage{bbold}
\usepackage[latin1]{inputenc}
\usepackage[T1]{fontenc}

\usepackage{hyperref}
\usepackage{times}
\usepackage{amsmath,amsfonts,amssymb}
\usepackage{graphicx,graphics}
\usepackage{diagbox}
\usepackage{epstopdf,csquotes,}
\usepackage{color,textcomp,bm}
\usepackage{subfigure,array,xspace} 
\usepackage[squaren,Gray,cdot]{SIunits}
\usepackage{float}
\usepackage{mathdots}
\usepackage{MnSymbol}
\usepackage{placeins}
\usepackage{multibib}

\usepackage{xr}
\newcommand*{\rev}[1]{\textcolor{black}{#1}}

\newcommand*{\alex}[1]{\textcolor{black}{#1}}
\newcommand{\inp}{_{\textrm{in}}}
\newcommand{\out}{_{\textrm{out}}}

\begin{document}

\begin{abstract}
Volcanic eruptions necessitate precise monitoring of magma pressure and inflation for improved forecasting. Understanding deep magma storage is crucial for hazard assessment, yet imaging these systems is challenging due to complex heterogeneities that disrupt standard seismic migration techniques. Here we map the magmatic and hydrothermal system of the La Soufrière volcano in Guadeloupe by analyzing seismic noise data from a sparse geophone array under a matrix formalism. Seismic noise interferometry provides a reflection matrix containing the signature of echoes from deep heterogeneities. Using wave correlations resistant to disorder, matrix imaging successfully unscrambles wave distortions, revealing La Soufrière's internal structure down to 10 kilometers with 100-meter resolution. This method surpasses the diffraction limit imposed by geophone array aperture, providing crucial data for modeling and high-resolution monitoring. We see matrix imaging as a revolutionary tool for understanding volcanic systems and enhancing observatories' abilities to monitor dynamics and forecast eruptions.
\end{abstract}

\title{Matrix imaging as a tool for high-resolution monitoring of deep volcanic plumbing systems with seismic noise}

\author         {Elsa Giraudat}
\affiliation    {Institut Langevin, ESPCI Paris, PSL University, CNRS, Paris, France}
\author         {Arnaud Burtin}
\affiliation    {Institut de Physique du Globe de Paris, {Universit\'{e} Paris Cit\'{e}, CNRS,} Paris, France\\
$^*$Corresponding author (alexandre.aubry@espci.fr)}
\author         {Arthur Le Ber}
\affiliation    {Institut Langevin, ESPCI Paris, PSL University, CNRS, Paris, France}
\author         {Mathias~Fink}
\affiliation    {Institut Langevin, ESPCI Paris, PSL University, CNRS, Paris, France}
\author         {Jean-Christophe Komorowski}
\affiliation    {Institut de Physique du Globe de Paris, {Universit\'{e} Paris Cit\'{e}, CNRS,} Paris, France\\
$^*$Corresponding author (alexandre.aubry@espci.fr)}
\author         {Alexandre Aubry$^{*,}$}
\affiliation    {Institut Langevin, ESPCI Paris, PSL University, CNRS, Paris, France}

\date{\today}
   \maketitle

   \clearpage 

\noindent {\Large \textbf{Introduction}} 

In everyday life, a multitude of sensors surround us to monitor our environment. In wave physics, those sensors can be active and work together to control the wave-field at will whether it be for focusing~\cite{Mosk2012} or communication~\cite{Moustakas2000} purposes. For imaging, the problem is often ill-posed because of the medium complexity and/or the sensor array sparsity. This is particularly the case in seismology, where the topography of the site under investigation can be so irregular that it is \rev{prohibitive} to deploy a large and dense network of geophones.

This paper addresses the issue of seismic imaging in complex areas such as volcanoes or fault zones based on data recorded by a sparse array of seismometers. The goal is to provide high spatial resolution and in-depth imaging of such critical areas that are of paramount importance for Earth sciences. To that aim, we will build on a matrix imaging approach imported from other fields than geophysics, such as medical ultrasonics~\cite{Chau2019,lambert_distortion_2020} and optical microscopy~\cite{Kang2017,badon_distortion_2020} that were designed for scales ranging from a few centimeters for ultrasonic waves to a few hundreds of nanometers for light. In contrast with concurrent seismic methods such as full waveform inversion~\cite{Virieux2009}, the strength of matrix imaging lies in the fact that: (i) it does not rely on a sophisticated wave velocity model whose knowledge is often limited and uncertain in geophysics; (ii) it is robust with respect to data quality which is a frequent issue in seismology.

Matrix imaging relies on the array response matrix that contains the set of impulse responses between each seismometer. Although a geophone is purely passive, cross-correlation of seismic noise received at two stations is known to converge toward the Green's function between receiving stations~\cite{Weaver2001,campillo_long-range_2003}, as if one of them had been used as source, thus paving the way to passive matrix imaging~\cite{Blondel2018,touma_distortion_2021,Touma2023}. As surface waves dominate ambient noise, most past studies on the topic aimed at extracting surface wave properties from ambient noise correlation functions \rev{(NCFs)}~\cite{Shapiro2005}. However, they also contain the contribution of body waves reflected by deep structures~\cite{roux_p-waves_2005} {and fluid reservoirs~\cite{Blondel2018}.} 

As a proof-of-concept, we here exploit seismic noise recorded by a sparse geophone network deployed at the surface of the La Soufrière volcano \alex{of Guadeloupe}~\cite{IDPDGDP2021,Burtin2017}. The covariance matrix of this seismic noise provides the reflection matrix that contains all the available information on the underground reflectivity. {A numerical focusing process, often referred to as \textit{redatuming} in seismology~\cite{Schuster2008}, can then be applied to provide a confocal image of the subsoil reflectivity~\cite{Blondel2018}. This image in reflection is directly proportional to the axial fluctuations of the acoustic impedance associated with length scales smaller than the wavelength~\cite{Sentenac2018}. It is therefore an extremely relevant observable for highlighting the presence of fluid-rock interfaces. However, the quality of the confocal image} is drastically degraded by: (i) the mismatch between the wave velocity model and its true distribution that gives rise to a foggy image; (ii) the sparsity and finite size of the geophone network that limit its resolution. The former problem is {solved} by compensating wave distortions from the Earth surface, thereby revealing an helical \alex{conduit} in the {upper} part of the volcano. The sparsity issue is addressed by an iterative phase reversal process driven from the $\mathbf{k}-$space~\cite{Najar2023,Bureau2023} that resolves the deep reflectors with a transverse resolution of the order of a half-wavelength ($\sim$100 m), thereby breaking the free space diffraction-limit usually limited by the array aperture. \rev{This is in contrast with a previous study on the Erebus volcano~\cite{Blondel2018} that provided a reflectivity image of its main structures (lava lake, magma chambers) but at a poor lateral resolution; no compensation for wave distortions was performed.} {\rev{Here}, the inner structure of the volcano is revealed up to a depth of 10 km. It shows sub-horizontal bodies linked by thinner sub-vertical structures that match the current state-of-the-art conceptual and data-derived view of transcrustal magmatic systems. Such a mush-based model applies to numerous volcanic systems and has indeed been theorized for La Soufrière of Guadeloupe~\cite{Moretti2020,Metcalfe2021,Metcalfe2022}.}\\

\noindent {\Large \textbf{Results}}\\

\noindent {\large \textbf{Canonical Reflection Matrix}}

\begin{figure*}
\centerline{\includegraphics[width=\textwidth]{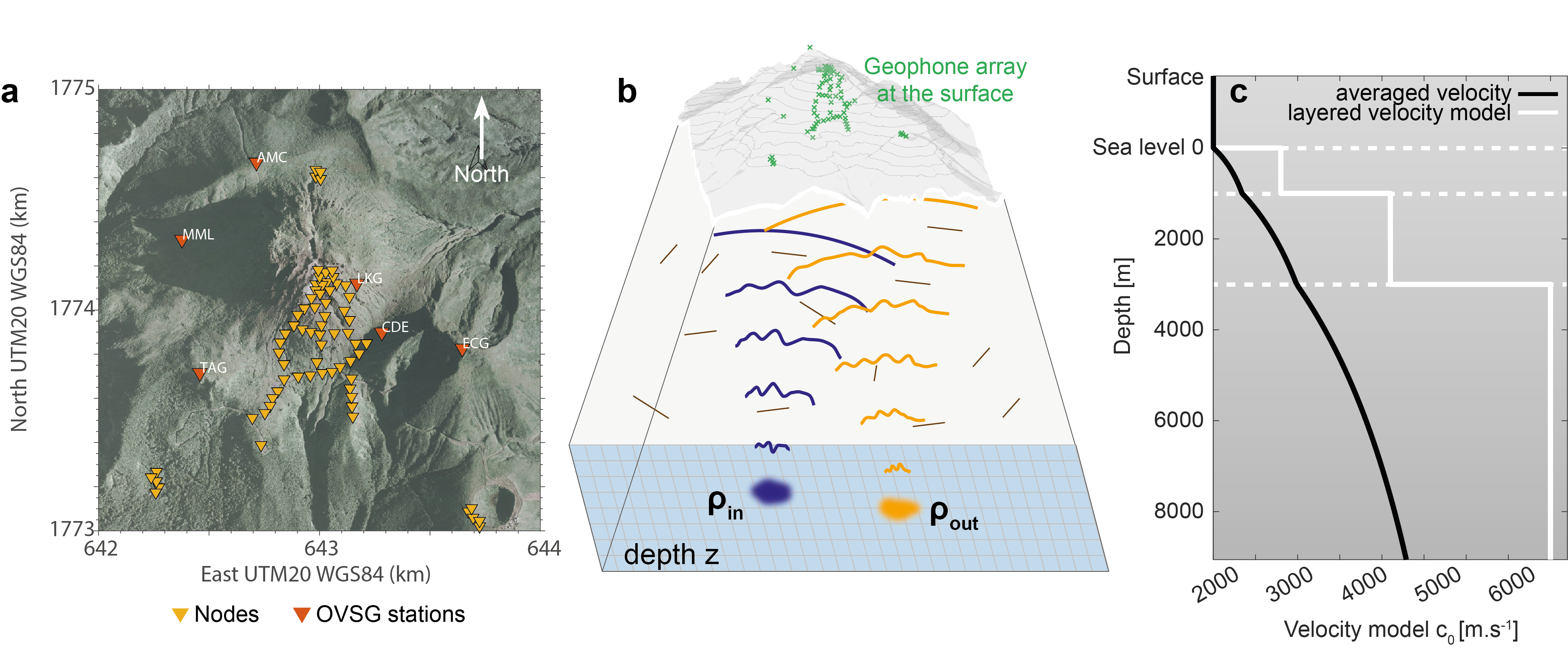}}
\caption{\textbf{Passive imaging of La Soufrière volcano.} \textbf{a}, Map of the 76 geophones installed above La Soufrière (Guadeloupe, France). {Both permanent stations (red) and temporary nodal array (\rev{orange}) are used}. Map data: Google, CNES, Airbus, 2023. \textbf{b}, Covariance matrix of seismic noise acquired during 2 months is post-processed to obtain the impulse responses between a set of virtual geophones {identified by their position $\mathbf{r}_{\textrm{in/out}}=(\bm{\rho}_{\textrm{in/out}},z)$ and} mapping the inside of the volcano. \textbf{c}, 1D-velocity model{~\cite{Dorel1979}} used for the seismic data redatuming process.}  
\label{figA_network_BF_c0}
\end{figure*}

Figure~\ref{figA_network_BF_c0}a shows the {virtual} network of 76 geophones whose distribution has been dictated by the topography of the volcano. {It} spans over a lateral extension $d_{||}=1300$~m and a vertical range $d_z= 500$ m. 
{The impulse response \rev{$R (\mathbf{g}_i,\mathbf{g}_j,t)$} between each pair of stations $(i,j)$ \rev{(located at positions $\mathbf{g}_i$ and $\mathbf{g}_j$, respectively)} is estimated by cross-correlation of ambient seismic noise \cite{campillo_long-range_2003} (Methods). \rev{To avoid the detrimental effect of fumaroles which are extremely localized noise sources (Supplementary Note 2), only the anti-causal component of the NCFs has been considered to estimate the impulse reponse between each geophone (Supplementary Note 3).} The set of impulse responses is stored in a time-dependent response matrix $\mathbf{R} (t)$.}

This {canonical reflection matrix} is powerful since it enables {a post-processing projection of} seismic data into different mathematical bases. {The} reflection matrix can {be} investigated into the plane wave basis (or $\mathbf{k}$-space) or any plane in the real space that sits between the Earth surface ($\mathbf{u}$) and the expected focal plane ({$\bm{\rho}$}) at a given time-of-flight $t$. To project the seismic data in these latter bases, a wave velocity model is nevertheless required.\\

\noindent {\large \textbf{Wave velocity Modeling}}

As we consider only the vertical component of the impulse responses, collected echoes are assumed to be mainly associated with P-waves~\cite{roux_p-waves_2005}. Thus, we adopt in the following a homogeneous P-wave velocity model. More precisely, for each depth, we define a homogeneous velocity distribution whose value is calculated on the basis of the four-layer large scale velocity model{~\cite{Dorel1979}} (Fig.~\ref{figA_network_BF_c0}c). This value ranges from $c_0 = 2000$ m.s\textsuperscript{-1} at shallow depth to $c_0 = 4300$ m.s\textsuperscript{-1} at depth $z= 10$ km below the surface. The detailed evolution of the wave velocity model $c_0(z)$ with respect to depth is given in Fig.~\ref{figA_network_BF_c0}c. The assumed background velocity model is rough but is, as we will see, sufficient to image the volcano by leveraging the matrix approach. \\

\noindent {\large \textbf{Confocal Redatuming}}
 
In a first step, the velocity model is used to back-propagate in-depth the recorded echoes gathered in the canonical reflection matrix $\mathbf{R}$ in order to retrieve local reflectivity information at each depth of interest. Back-propagation is commonly achieved by applying appropriate time delays at emission and at reception to migrate echoes in post-processing. Such focusing operations are frequently used in imaging and are in particular known as redatuming in seismic exploration~\cite{berkhout_unified_1993}.  The matrix formalism offers a convenient framework to easily perform such beamforming in post-processing, especially in the frequency domain where these operations are described using simple matrix products~\cite{Blondel2018,lambert_reflection_2020,touma_distortion_2021,Touma2023} {(Methods, \rev{Eq.~\ref{RggToRrrcoeff}})}. 
 
\begin{figure*}
\centerline{\includegraphics[width=0.7\textwidth]{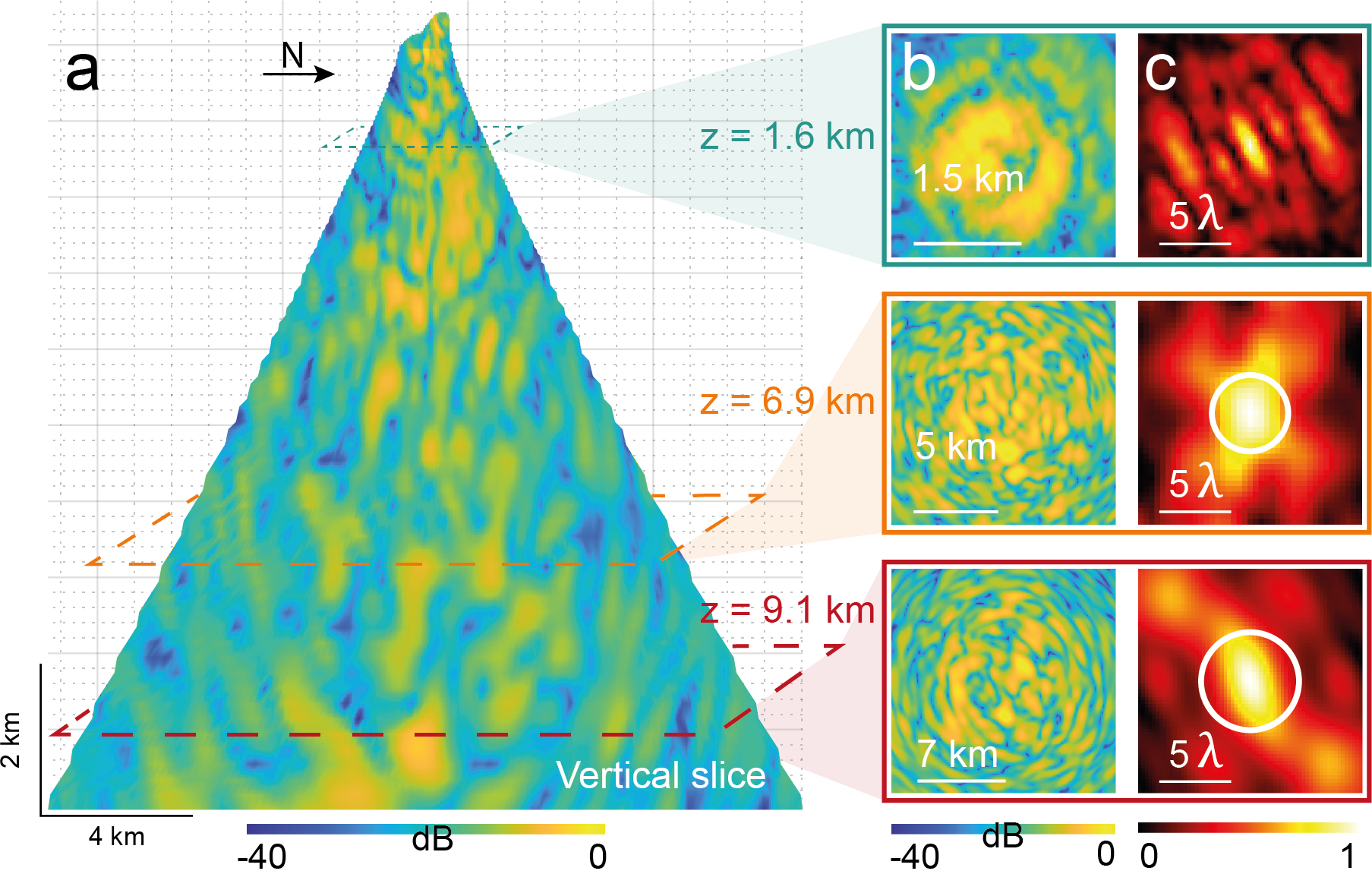}}
\caption{\textbf{Confocal redatuming}. \textbf{a}, Vertical slice of the 3D confocal image along the South-North direction. {This image is shown after depth compensation of seismic wave attenuation (section S5).} \textbf{b}-\textbf{c}, Horizontal slices (\textbf{b}) and associated \rev{reflection point spread functions} (\textbf{c}) at depths $z=1.6$ km, {$6.9$ km and $9.1$} km below the summit. The spatial extension $\delta \rho_u$ (Eq.~\ref{resolution}) of the theoretical diffraction-limited focal spot is denoted as a white circle.}  
\label{figOrig}
\end{figure*}

The result is a focused reflection matrix {$\mathbf{R}_{\bm{\rho\rho}}(z) = [R(\bm{\rho}_{\textrm{out}},\bm{\rho}_{\textrm{in}},z)]$ at each depth $z$ \rev{(Methods, Eq.~\ref{Rbdband})}} that contains the inter-element impulse responses between a set of virtual sources at {$\mathbf{r}_{\textrm{in}}=(\bm{\rho}_{\textrm{in}},z)$} and virtual receivers at {$\mathbf{r}_{\textrm{out}}=(\bm{\rho}_{\textrm{out}},z)$} mapping the inner structure of the volcano (Fig.~\ref{figA_network_BF_c0}b). \rev{Note that those responses are time-gated around the expected ballistic time ($t\sim 2z/c_0$) and thus frequency-averaged over the whole bandwidth. The diagonal elements of $\mathbf{R}_{\bm{\rho\rho}}(z)$} are associated with coincident \textit{input} and \textit{output} focusing points {($\bm{\rho}_{\textrm{in}} = \bm{\rho}_{\textrm{out}}$, see Methods)}. {After compensation of wave attenuation with depth \rev{(Methods, Eq.~\ref{norma})}, a 3D confocal image of the volcano is obtained (Fig.~\ref{figOrig}a) with horizontal cross-sections shown for different depths in Fig.~\ref{figOrig}b : (\textit{i})~$z=1.6$~km \textit{i.e} where the most abundant seismicity occurs at La Soufrière; (\textit{ii})~$z=6.9$~km \textit{i.e} at the level of the magma reservoir whose depth range is expected between 5.6 and 8.5 km~\cite{Pichavant2018}; (\textit{iii})~$z=9.1$~km \textit{i.e} beyond the magma reservoir.}
  
Whether it be on the transverse or the vertical view (see Supplementary Movies 1, 2 and 3), some scattering structures seem to emerge at different locations {in Figs.~\ref{figOrig}a,b} but the overall structure appears to be fully blurred, suggesting a high level of aberrations. Such a raw confocal image is indeed very sensitive to aberrations and its interpretation should be extremely cautious.\\

\noindent {\large \textbf{Assessing Focusing Quality}}

The focusing quality can actually be assessed by considering the off-diagonal elements of {$\mathbf{R}_{\bm{\rho\rho}}(z)$ ($\bm{\rho}_{\textrm{in}} \neq \bm{\rho}_{\textrm{out}} $)} that provide an estimator of the point spread function in reflection (RPSF) as a function of the relative position {$\Delta \bm{\rho}=\bm{\rho}_{\textrm{out}}-\bm{\rho}_{\textrm{in}}$}~(Methods). Figure~\ref{figOrig}c displays the evolution of RPSF for different depths and highlights a drastic spreading of the back-scattered energy over off-diagonal elements of {$\mathbf{R}_{\bm{\rho\rho}}(z)$}. This is a direct manifestation of the gap between the wave velocity model and its true distribution in the volcano. \rev{Note that an imperfect convergence of the NCFs towards the Green's functions can also lead to an additional incoherent background on the RPSF (Supplementary Note 1). At this stage, it is difficult to discriminate between these different phenomena but we will see that the roughness of the wave velocity model is the main issue in the present case.} 

In absence of aberration, all the back-scattered energy would be contained in a \rev{diffraction}-limited confocal spot (white circle in Figure \ref{figOrig}c) whose size is governed by the angle $\theta_u=\tan^{-1}\left( d_{||}/2z\right)$ under which the geophone array is seen by the focusing point:  
\begin{equation}
\label{resolution}
\delta \rho_u = \lambda / \left( 2 \sin \theta_u \right).
\end{equation}
In Fig.~\ref{figOrig}c, the focused wave-field spans over a much larger area than this ideal focal spot and strong side lobes appear around the main central lobe, indicating that images suffer from a high level of aberration. \\

\noindent {\large \textbf{Overcoming Aberrations}}
\begin{figure*}    
\centerline{\includegraphics[width=\textwidth]{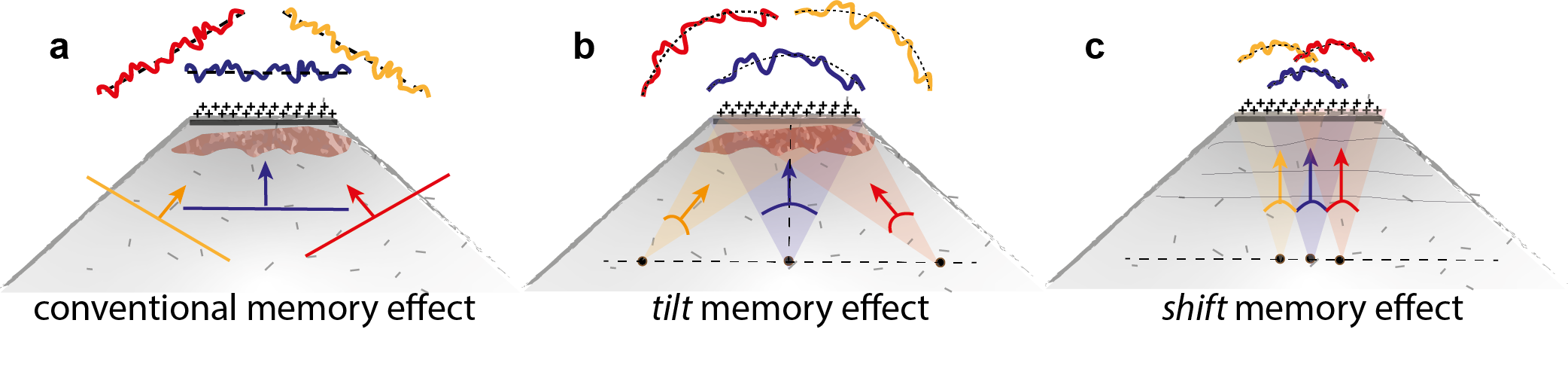}}
\caption{\rev{\textbf{Memory effect}. \textbf{a}, Conventional memory effect:  A plane wave coming from depth gives rise to a distorted wave-front at the Earth surface due to wave velocity heterogeneities lying close to the Earth surface (aberrating layer). If the incident plane wave direction is tilted, the same transmitted wave-field is obtained but tilted by the same angle as the incident plane wave. \textbf{b}, {Tilt} memory effect: If we now consider point-like sources at a given depth, an angular memory effect can be observed through a thin aberrating layer: The transmitted wave-fronts display similar wave distortions but are tilted with respect to each other by an amount dictated by the position of point-like sources. \textbf{c}, {Shift} memory effect: In a multi-layered medium, the transmitted wave-fronts display similar wave distortions but, this time, laterally shifted with respect to each other.} }
\label{memory}
\end{figure*}

\begin{figure*}    
\centerline{\includegraphics[width=\textwidth]{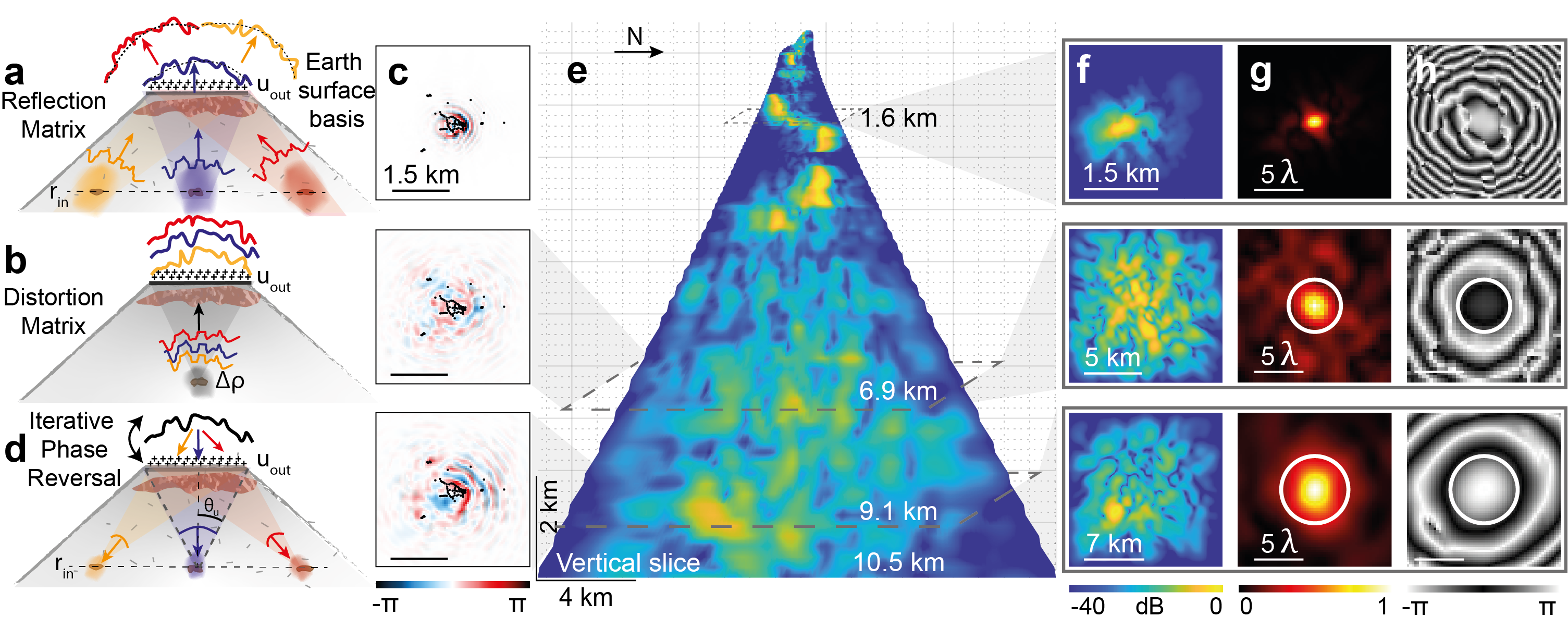}}
\caption{\textbf{Exploiting memory effect for overcoming aberrations}. \textbf{a}, Schematic view of back-scattered wave-fronts in the Earth surface basis ($\mathbf{u}$) generated by virtual sources ($\mathbf{r}_{\textrm{in}}$) at a given depth $z$. \textbf{b}, The extraction of wave-distortions amounts to an angular de-scan of each input focal spot. \textbf{c}, Iterative phase reversal applied to the $\mathbf{D}-$matrix at each depth provides an aberration transmittance whose phase is here shown at depths $z=1.6$ km, {$6.9$ km and $9.1$} km and whose modulus is encoded as a transparency mask. Black dots indicate the lateral position of geophones. \textbf{d}, Phase conjugation and tilt of such aberration phase laws enable an adaptive focusing process on each point of the subsoil. \textbf{e}, Vertical slice of the 3-D confocal image obtained by means of these optimized focusing laws. {This image is shown after a depth compensation of seismic wave attenuation (section S5).} \textbf{f}, Corresponding horizontal slices at depths $z=1.6$ km, {$6.9$ km and $9.1$} km. \textbf{g}, Modulus and \textbf{h}, phase of the resulting RPSFs at the same depths. The spatial extension $\delta \rho_u$ (Eq.~\ref{resolution}) of the theoretical diffraction-limited focal spot is denoted as a white circle.} 
\label{figCorrU}
\end{figure*}

\rev{To isolate and compensate for these aberration effects, we build upon a physical phenomenon referred to as the memory effect in wave physics~\cite{Osnabrugge2017}.The memory effect ensures that a pattern of random phase shifts imparted to a plane wave-front by an aberrating layer (wave velocity heterogeneity) keeps its general ``shape'' but is ``tilted'' if the orientation of the incoming plane wave is also tilted (Fig.~\ref{memory}a). This implies that the distorted wave-front coming from a point-like source inside the medium will be tilted if the source is laterally shifted (Fig.~\ref{memory}b). In the present case, we do not have  any real source inside the volcano but virtual sources produced by the redatuming process. In a similar way, a shift of the virtual source will imply a tilt of the reflected wave-front at the Earth surface as sketched by Fig.~\ref{figCorrU}a.}

To exploit this \rev{{tilt}} memory effect, our strategy is thus the following (Methods): (\textit{i}) project the reflection matrix between the focused basis $(\mathbf{r})$ and the Earth surface basis $(\mathbf{u})$ (Fig.~\ref{figCorrU}a); (\textit{ii}) highlight the angular correlations of the reflected wave-field by building a dual-basis matrix (the distortion matrix $\mathbf{D}$) that connects any input focal point in the medium with the distortion exhibited at the Earth surface by the corresponding reflected wavefront (Fig.~\ref{figCorrU}b); (\textit{iii}) take advantage of the angular correlations between those wave distortions to accurately estimate the aberration phase transmittance in the Earth surface basis through an iterative phase reversal algorithm (Fig.~\ref{figCorrU}c); (\textit{iv}) phase conjugate the resulting transmittance to tailor adaptive focusing laws that shall compensate for the volcano{'s} heterogeneities (Fig.~\ref{figCorrU}d).


Vertical and horizontal cross-sections of the resulting confocal image are displayed in Figs.~\ref{figCorrU}e and f, respectively. The comparison with the initial image demonstrates the benefit of the correction process, especially at shallow depths ($z<4$ km) where the twisted {conduit} of the volcano is revealed. The comparison of the original and the corrected RPSFs (Figs.~\ref{figOrig}c and \ref{figCorrU}g) confirms that the focusing quality is significantly improved in this depth range: Whereas the original RPSF (top panel in Fig.~\ref{figOrig}b) spreads far beyond the diffraction-limited focal spot, the transverse extension of the corrected RPSF is drastically reduced.  However, the gain in image and focusing quality is more modest at larger depths (Figs.~\ref{figCorrU}e,f). The RPSFs still exhibit secondary lobes, a manifestation of residual aberrations (Fig.~\ref{figCorrU}g). Moreover, the spatial extension of the central lobe is limited by the geophone network aperture (Eq.~\ref{resolution}). As a consequence, the deep plumbing system of the volcano{, in particular \alex{the deepest regions of the transcrustal magmatic system and its magma storage zones beyond 5 km depth},} cannot be resolved.\\

\noindent {\large \textbf{Beating Diffraction}}

\begin{figure*}    
\centerline{\includegraphics[width=\textwidth]{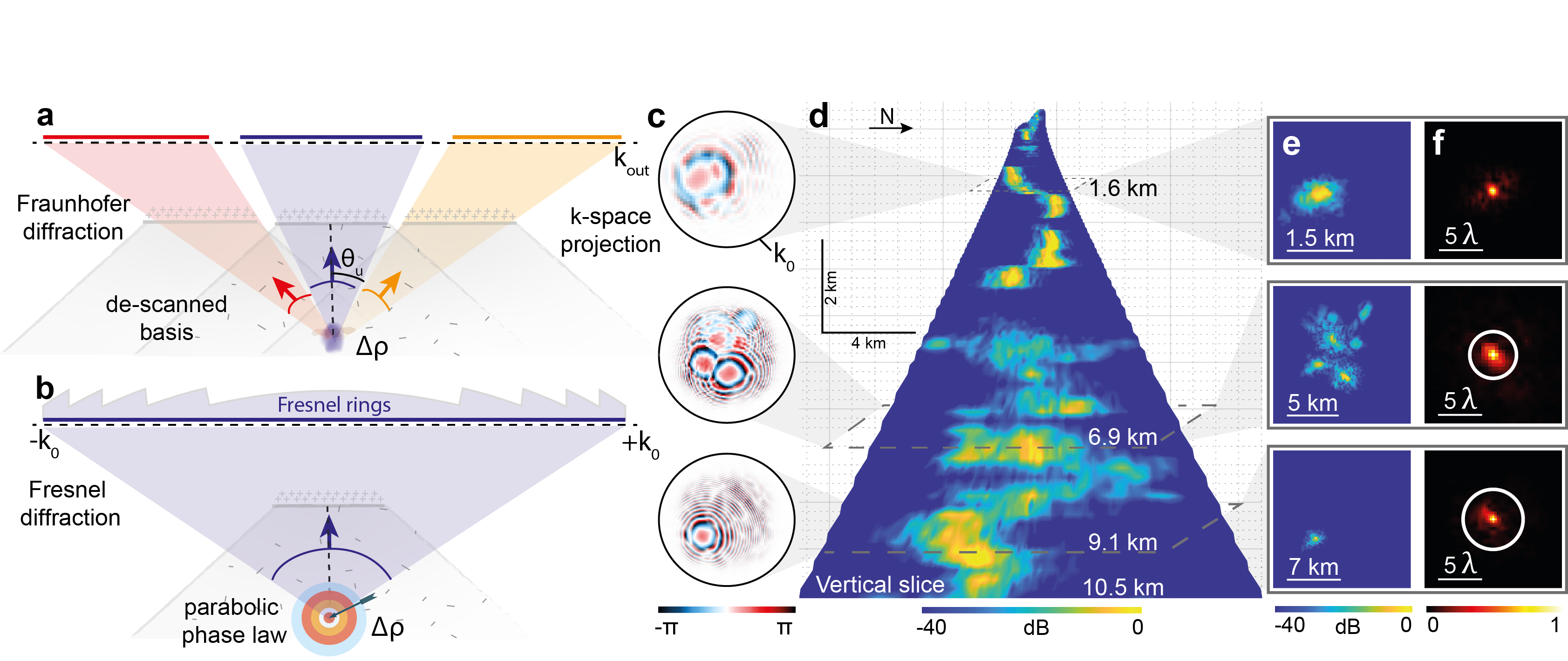}}
\caption{\textbf{Overcoming diffraction by operating the focusing process from the $\mathbf{k}-$space}.
\textbf{a}, The distorted wave-field in the $\mathbf{k}-$space amounts to a lateral de-scan of each reflected echoes. Under the Franhofer approximation, the far-field projection of each focal spot is limited by the de-scanned geophone network aperture. \textbf{b}, In the present case, Fresnel diffraction gives rise to a modulation of each focal spot by a parabolic phase law. The diffraction pattern of each reflector then corresponds to Fresnel rings that cover the whole diffraction disk of radius $k_0$. For sake of clarity, only the contribution of a central reflector is displayed. \textbf{c}, Iterative phase reversal applied to $\mathbf{D}$ extracts those diffraction patterns here shown at depths $z=1.6$ km, {$6.9$ km and $9.1$ km}. \textbf{d}, Vertical slice of the resulting image (same view as in Figs.~\ref{figOrig}a and \ref{figCorrU}d). {This image is shown after depth compensation of seismic wave attenuation \rev{(Methods, Eq.~\ref{norma})}.} \textbf{e}-\textbf{f}, Corresponding horizontal slices (\textbf{e}) and RPSFs (\textbf{f}) at the same depths as in (\textbf{c}). The spatial extension $\delta \rho_u$ (Eq.~\ref{resolution}) of the aperture-limited focal spot is denoted as a white circle.}  
\label{figCorrK}
\end{figure*}
{Strikingly}, an analysis of wave distortions from the $\mathbf{k}$-space will allow us to break this fundamental limit. In {the plane wave} basis, each distorted wave-field corresponds to the diffracted patterns of each laterally de-scanned output focal spot~\cite{lambert_distortion_2020}. In a far-field approximation, the contribution of each scatterer would emerge onto limited parts of the $\mathbf{k}-$space because of the finite size of the geophone array (see Fig.~\ref{figCorrK}a and Supplementary Note 9). However, the focal spots also exhibit a parabolic phase law scaling as $\exp (jk_0 |\Delta \bm{\rho}|^2/z)$\rev{, resulting} from the curvature of focused wave-fronts  (see Fig.~\ref{figCorrU}h and Methods). Projected in the $\mathbf{k}$-space, the associated transfer function is thus a superposition of Fresnel rings associated with each reflector. The support of those Fresnel rings is not limited by the geophone network aperture ($k_0 \sin \theta_u$) but covers the whole diffraction disk of radius $k_0$ (Fig.~\ref{figCorrK}b). Iterative phase reversal applied to the $\mathbf{D}-$matrix expressed in the $\mathbf{k}-$space leads to a focusing law (Fig.~\ref{figCorrK}c) that realigns the phase of each spatial frequency component such that the focal spot size reduces to the diffraction limit $\delta \rho_0 \sim \lambda/2$ (Supplementary Note 10). It leads to a new confocal image whose several cross-sections are displayed in Figs.~\ref{figCorrK}d and e. In particular, a complex multi-lens melt reservoir is revealed by Fig.~\ref{figCorrK}d beyond a depth of 5 km. The comparison with the previous image (Figs.~\ref{figCorrU}e and f) highlights the spectacular gain in terms of contrast and resolution provided by a $\mathbf{k}-$space analysis of the $\mathbf{D}-$matrix. This observation is confirmed by the new RPSFs displayed in Fig.~\ref{figCorrK}f. Compared to their previous version (Fig.~\ref{figCorrU}g), the diffuse background has been suppressed by a compensation of residual \rev{axial} wave distortions~\cite{Lambert2022} exhibiting a \rev{{shift}} memory effect~\cite{Osnabrugge2017} \rev{(Fig.~\ref{memory}c)}. More importantly, the RPSF extension is now of the order of $\lambda/2\sim 100$ m over the whole considered depth range, thereby beating the usual aperture-limited resolution (Eq.~\ref{resolution}) displayed by conventional imaging methods.

One necessary condition for this striking performance is the sparsity of the volcano reflectivity with only a few reflectors emerging at each depth (Fig.~\ref{figCorrK}e). As the signature of each reflector is independent, we are able to focus simultaneously on each scatterer provided that they are not too numerous. More precisely, the contrast of the confocal image will typically scale as the ratio between the number of independent geophones and the number of reflectors lying at each depth. \rev{This explanation shed a new light onto the high-resolution image of the San Jacinto Fault Zone provided by matrix imaging in a previous study~\cite{touma_distortion_2021}. While super-resolution was initially accounted for by physical phenomena such as multiple scattering or wave channeling, medium sparsity and wave-front curvatures most likely explain the striking performance of matrix imaging in that seismic configuration as well. The phenomenon described above is therefore not just a curiosity of La Soufrière; it can actually be exploited in many other seismic situations.} 
\vspace{5 mm}



\noindent {\large \textbf{Unveiling the plumbing system of La Soufrière}}

\begin{figure*}    
\centerline{\includegraphics[width=\textwidth]{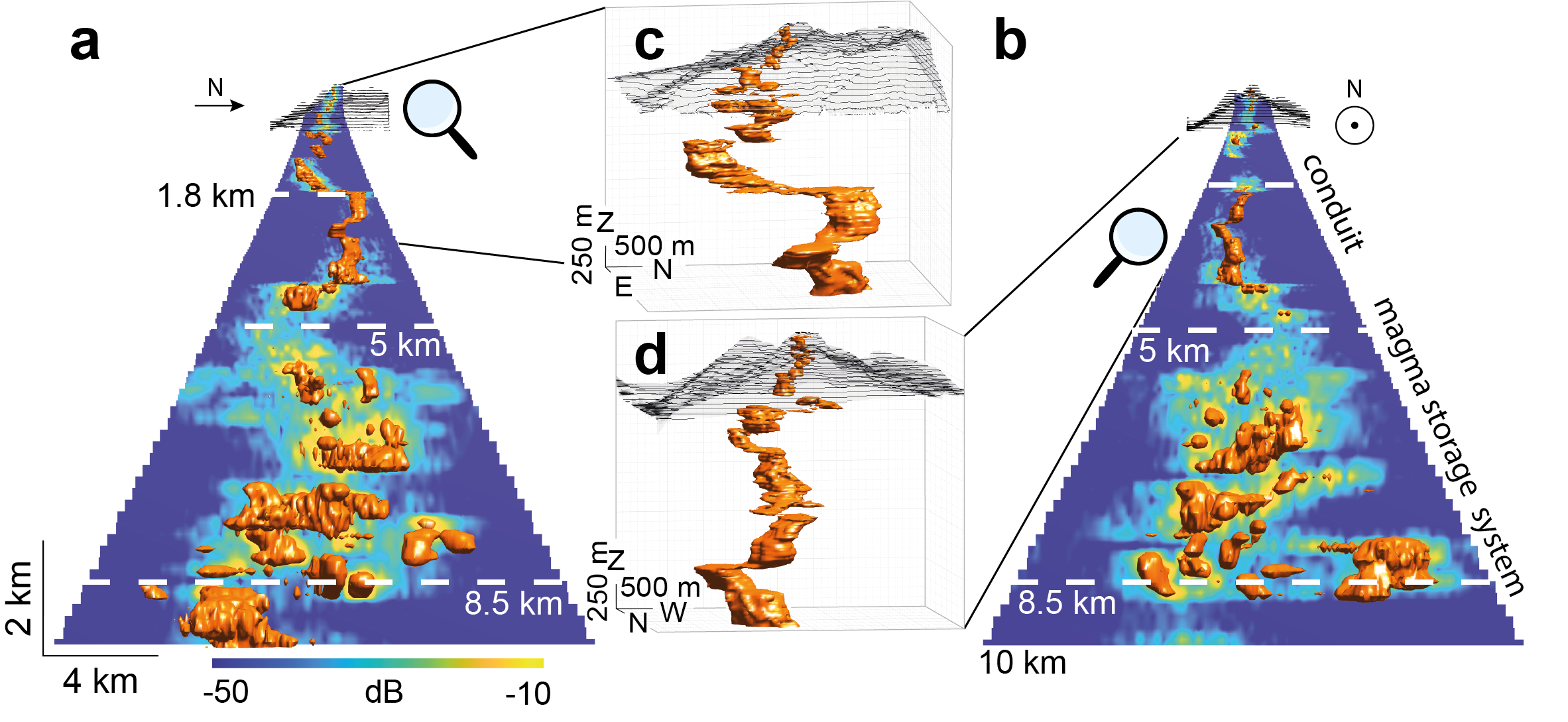}}
\caption{\textbf{Three-dimensional view of {the first 10 km of the hydrothermal and magmatic system of La Soufrière}}. \textbf{a}-\textbf{b}, Isosurface plots of the three-dimensional image of the volcano viewed from East and North, respectively. The isosurface is fixed to be {-15 dB}. \textbf{c}-\textbf{d}, Corresponding zooms on the first $3$ km depth. The isosurface is fixed to be {-10 dB}. {This image is shown after a depth compensation of seismic wave attenuation (\rev{Methods, Eq.~\ref{norma}}).}}  
\label{FigIso}
\end{figure*}

Figure \ref{FigIso} shows two perpendicular views of La Soufrière down to a depth of \alex{10.5} km below the summit (see also Supplementary Movies 1, 2 and 3). {Based on an analysis of the P-wave reflected wave-field, it displays} the iso-surfaces of the confocal image obtained at the end of the matrix imaging process. As outlined above, the upper part of the volcano{, from a depth of 5 km up to the surface,} exhibits the clear signature of a tortuous {conduit} that finds its way through the \rev{host rock} forming the {upper part of the} volcano. On the contrary, its deep structure{, between \textit{ca.} 5 and 8.5 km depth,} induces a more diffuse scattering {that is compatible with the existence of a vertical succession of several sub-horizontal and irregular globular coalescing structures. Those elements are superimposed over a distance of a few kilometers and linked together by narrow sub-vertical diffuse structures. The sub-horizontal structures extend laterally over a distance of about 8 km. \rev{We interpret them as sub-horizontal lenses containing an unknown volume of potentially eruptive magma.}. The presence of the \alex{superimposed magma storage zones} is also highlighted by the depth-dependence of unnormalized scattering signal displayed in Supplementary Figure~12. \alex{The magma strorage system ($z=$5-8.5 km) exhibits a weaker reflectivity probably due the presence of extended magma volumes. The enhancement of the confocal signal above the outer carapace of the magma storage zone ($z=$ 3.5-5 km) may be induced by gases and/or liquid and/or supercritical hydrothermal and magmatic fluids that are present in the pores of the \rev{host rock} along special zones of elevated porosity-permeability. The increase of reflectivity observed at the bottom of the magma storage system ($z=$ 8.5-10 km) is probably due to a strong back-reflection at the eruptible melt / \rev{host rock} interface.} The 3D-view of the internal structure of La Soufrière volcano displayed in Fig.~\ref{FigIso} thus constitutes a remarkable advance beyond the current state-of-the-art because it confirms, for the first time, with great detail and striking similarity the typical structure of transcrustal magmatic systems below volcanoes that has been predicted by previous conceptual and petrological models~\cite{Cashman2017,Cassidy2018,Edmonds2019}} \rev{and that can be observed in the field in old dissected magmatic systems.}

{Transcrustal magmatic systems consist of vertically-arranged piles of lenses of magmatic mushes more or less ductile (intricate network of crystals and intersticial melt fraction), eruptible melt, and magmatic fluids that extend laterally. This model of a magmatic plumbing system has been described at many other volcanoes~\cite{Combier2015,Edmonds2016,Koulakov2018,Winslow2022}. The internal image of the volcano revealed by Fig.~\ref{FigIso} strikingly matches the complex structure described by recent studies on La Soufrière of Guadeloupe~\cite{Metcalfe2021,Metcalfe2022}.}

{Last but not least, the  seismic confocal image of La Soufrière shows that the main magmatic plumbing systems extends from about 5 km below the surface to a depth of at about 8.5 km, values in agreement with those determined by independent petrological studies~\cite{Pichavant2018}  who showed that, for the last magmatic eruption of La Soufrière in 1530 CE, the top of the magma storage zone was located between 5.6 and 7 km and the base could not exceed 8.5 km.}

\vspace{5 mm}

 \noindent {\Large \textbf{Discussion}}

{Passive seismic matrix imaging 
reveals, for the first time, high-resolution features of the magma storage zone, its geometry and dimensions, its complex layered structure, its relative connectedness with other regions of the multi-layer transcrustal magmatic system, and the size and geometry of the upper final eruptive conduit.} {The impedance contrast in this complex image also offers the potential, upon further analysis, to distinguish zones of mush from those of eruptible melt, their relative volume, their position in the system. Hence, it can lead to the estimation of parameters such as pressure, temperature, volatile saturation, density contrast, and the connectivity to the surface in evolving magmatic systems, parameters that drive volcanic eruptions. }

{The strength of this new imaging method lies in its robustness with respect to sparsity of the geophone array, inaccuracy of the wave velocity model, \rev{presence of spurious arrivals in the NCFs (Supplementary Note 4) and their imperfect convergence towards the Green's functions (Supplementary Note 1). Despite the seemingly random feature of the computed NCFs (Supplementary Note 3), matrix imaging provides a coherent image of the internal structure of the volcano at an unprecedented resolution to the best of our knowledge. This robustness with respect to data quality and migration model inaccuracy is a great asset with respect to more computationally intensive methods such as full waveform inversion}. 

In the future, \rev{matrix imaging} will be combined with time-lapse ability resulting from reiteration surveys at active unresting volcanoes and can be coupled with multi-parameter data analysis from other classic monitoring networks. Matrix imaging can therefore become a revolutionary game changer in the way scientists understand and model volcanic systems and how volcano observatories monitor their evolving dynamics to forecast their potential for hazardous eruptive activity that threatens the lives of 800 million people living within 100 km from a dangerous volcano~\cite{Papale2019}.}


\newpage

\noindent {\Large \textbf{Methods}} 

\noindent {\textbf{Acquisition of seismic data.}}

\noindent {The seismic data used in this study consists of a temporary nodal array of 65 geophones~\cite{Burtin2017} and 6 permanent stations~\cite{IDPDGDP2021} operated by the OVSG-IPGP (Volcanologic and Seismologic Observatory of Guadeloupe).} The geophone sensors were Zland 3C Gen2 (Fairfieldnodal) with a natural frequency of 5 Hz, recording at 500 samples per second and along 3 orthogonal directions (Vertical, North and East). The 6 OVSG seismic stations are 3 components broadband sensors, all having a flat response in the [1-50] Hz frequency band. The seismic records are sampled at 100 Hz. For this study only vertical components are used. The temporary nodal array was deployed from mid-November 2017 to mid-January 2018 during 2 sessions in order to download seismic data and recharge batteries. Since we moved the location of 5 geophones between both acquisition sessions, we ended with a virtual network of 76 sites (Fig.~\ref{figA_network_BF_c0}a), for which we can apply the computation of seismic NCFs.
\vspace{5 mm}

\noindent {\textbf{Noise correlation processing.}}  

\noindent The procedure to compute the seismic NCF mainly follows the stages detailed by Bensen \textit{et al.}~\cite{bensen2007}. Here, we summarize each step that we apply on seismic recordings whether it was a temporary geophone or a permanent seismic sensor. (1) We detrend each hourly vertical seismic record and removed the mean. (2) We remove the instrument response to homogenize the seismic signals and we applied a band-pass filter between 1 Hz and 20 Hz. (3) We resample the seismic record to a unique sample frequency of 100 Hz. (4) We apply a spectral and temporal normalization by proceeding to a spectral whitening followed by a 1-bit normalization to only keep the sign of the seismic signal. (5) We end with the computation of the NCF by cross-correlating hourly seismic records at each stations pair for time delays ranging from -30 to +30 seconds. To increase the signal to noise ratio of a NCF, we apply some quality checks and a waveform summation by first averaging the 24 hourly NCFs in a daily one, for which we discard hourly segments that were not coherent with the raw daily average (correlation coefficient threshold of 0.5). \rev{Finally, the} average over each daily NCF estimated during the 2 months of nodal array deployment\rev{, $\Gamma(\mathbf{g}_j,\mathbf{g}_i,t)$,} provides an estimation of the impulse response $R(\mathbf{g}_j,\mathbf{g}_i,t)$ between each couple of geophones $i$ and $j$ \rev{by only considering the anti-causal component to avoid the detrimental effect of fumaroles (Supplementary Note 3): $R(\mathbf{g}_j,\mathbf{g}_i,t)=\Gamma(\mathbf{g}_j,\mathbf{g}_i,-t)$ with $t$ spanning from 0 to +30 s.} The set of the estimated 2850 vertical impulse responses forms the canonical reflection matrix $\mathbf{R}_{\mathbf{g}\mathbf{g}}(t) = [R(\mathbf{g}_j,\mathbf{g}_i,t)]$.
\vspace{5 mm}

\noindent {\textbf{Broadband focused reflection matrix.}}  

\noindent To build the focused reflection matrix a temporal Fourier transform is first applied to $\mathbf{R}_{\mathbf{g}\mathbf{g}}(t)$ to get the set of monochromatic canonical reflection matrices $\overline{\mathbf{R}}_{\mathbf{g}\mathbf{g}}(f)$ over the desired frequency bandwidth $[10-20]$ Hz. The monochromatic matrices are then propagated at emission and at reception towards a focal plane at depth $z$ using the corresponding free-space Green propagator $\mathbf{G}_{0}(z,f)$ whose coefficients write: 
 \begin{equation}\label{G0gr}
     G_0(\bm{\rho},\mathbf{g},z,f) = \frac{e^{ -i 2 \pi f \sqrt{{\|\bm{\rho} - \mathbf{g}_{||}\|}^2 + {\|z- g_z\|}^2 }/c_0(z)}}{4\pi\sqrt{{\|\bm{\rho} - \mathbf{g_{||}} \|}^2 +  {\|z- {g_z}\|}^2}}.
 \end{equation}
$\mathbf{G}_{0}(z,f)$ describes the causal 3-D propagation of waves between any geophone $\mathbf{g} = (g_x,g_y,g_z) = (\mathbf{g_{||}},g_z)$ and any focusing point $\bm{\rho}= (x,y)$ in the focused basis at depth~$z$ in a supposed homogeneous medium with a wave velocity $c_0(z)$. The evolution of the wave velocity $c_0(z)$ with respect to depth is provided in Fig.~\ref{figA_network_BF_c0}c.  
Within the framework of matrix imaging, the projection of $\overline{\mathbf{R}}_{\mathbf{g}\mathbf{g}}(f)$ at each depth $z$ is described by the following matrix product: 
\begin{equation}\label{RggToRrr}  
     \overline{\mathbf{R}}_{\bm{\rho}\bm{\rho}}(z,f) =  \mathbf{G}^*_0(z,f)\times \overline{\mathbf{R}}_{\mathbf{g}\mathbf{g}}(f) \times \mathbf{G}_0^{\dag}(z,f)
 \end{equation}
or in terms of matrix coefficients : 
\begin{equation}
\label{RggToRrrcoeff}  
    \overline{ R}(\bm{\rho}\out,\bm{\rho}\inp,z,f) = \sum_{\mathbf{g}\out} G_0^*(\bm{\rho}\out,\mathbf{g}\out,z,f) ~ \sum_{\mathbf{g}\inp} \overline{R}(\mathbf{g}\out,\mathbf{g}\inp,f) ~  G_0^*(\bm{\rho}\inp,\mathbf{g}\inp,z,f)
 \end{equation}
where the symbols $*$, $\dag$ and $\times$ stand for phase conjugate, transpose conjugate and matrix product respectively. It leads to the set of monochromatic focused reflection matrices $\overline{\mathbf{R}}_{\bm{\rho}\bm{\rho}}(z,f) = [\overline{R}(\bm{\rho}\out,\bm{\rho}\inp,z,f)]$. Physically, each coefficient of $\overline{\mathbf{R}}_{\bm{\rho}\bm{\rho}}(z,f)$ contains the inter-element impulse response between a virtual source located at $\mathbf{r}_{\textrm{in}} =(\bm{\rho}_{\textrm{in}},z)$ and a virtual detector at $\mathbf{r}_{\textrm{out}}=(\bm{\rho}_{\textrm{out}},z)$ (see Fig.~\ref{figA_network_BF_c0}b).
 
 In order to enhance this axial resolution, a broadband focused reflection matrix $\mathbf{R}_{\bm{\rho}\bm{\rho}}$ can be derived at each depth by coherently summing the monochromatic matrices over the frequency bandwidth:
 \begin{equation}\label{Rbdband}
 {\mathbf{R}}_{\bm{\rho}\bm{\rho}}(z) = \int_{f^-}^{f^+} df ~ \overline{\mathbf{R}}_{\bm{\rho}\bm{\rho}}(z,f)
 \end{equation}
 with $f^\pm = f_0 \pm \Delta f/2$, $f_0 = 15 $ Hz, and  $\Delta f = 10 $ Hz. The operation amounts to a ballistic time gating of singly-scattered echoes at times $t\sim 2z/c_0(z)$. Thanks to this operation, the axial dimension of virtual geophones is greatly reduced and only limited by the frequency bandwidth $\Delta f$: $\delta z_0 \sim c_0(z) /\Delta f$. In the single scattering regime, the coefficients of $\mathbf{R}_{\bm{\rho}\bm{\rho}}(z)$ can be theoretically expressed as follows~\cite{lambert_reflection_2020}:
 \begin{equation}\label{RHH}  
     {R}(\bm{\rho}\out,\bm{\rho}\inp,z) = \int d\bm{\rho}{H}(\bm{\rho},\bm{\rho}_{\textrm{out}},z) {\gamma}(\bm{\rho},z) {H}(\bm{\rho},\bm{\rho}_{\textrm{in}},z)
 \end{equation}
where $\gamma(\bm{\rho},z)$ is the medium reflectivity at depth $z$. $H(\bm{\rho},\bm{\rho}_{\textrm{in/out}},z)$ corresponds to the point-spread-function (PSF), that is to say the spatial amplitude distribution of the focal spot around the focusing point $\mathbf{r}_{\textrm{in/out}}$. 
Its support defines the characteristic size of each virtual source at $\mathbf{r}_{\textrm{in}}=(\bm{\rho}_{\textrm{in}},z)$ and detector at $\mathbf{r}_{\textrm{out}}=(\bm{\rho}_{\textrm{out}},z)$. 

Under the Fresnel approximation, the transmit PSF $H$ is the product of a parabolic phase law that results from the curvature of focused wave-fronts and a focusing function~$F$ (Supplementary Note \rev{6}):
 \begin{equation}
 \label{fresnel2}
    H(\bm{\rho},\bm{\rho}_{\textrm{in/out}},z)= \rev{\frac{1}{z^2 }} \exp \left [-i \frac{k_0}{2z} \left (|\bm{\rho}|^2-|\bm{\rho}_{\textrm{in/out}}|^2 \right)\right] F   \left (\frac{\bm{\rho}-\bm{\rho}_{\textrm{in/out}}}{\lambda z}\right )
\end{equation}
The focusing function $F$ results from the convolution between the network PSF $O$ that accounts for diffraction and an aberration PSF $A$ that results from the mismatch between the wave velocity model and the true wave velocity distribution in the volcano (Supplementary Note \rev{6}):
 \begin{equation}
 \label{focusing_function}
    F   \left (\frac{\bm{\rho} -\bm{\rho}_{\textrm{in/out}}}{\lambda z}\right ) =O \otimes A \left (\frac{\bm{\rho} - \bm{\rho}_{\textrm{in/out}}}{\lambda z} , z \right )
\end{equation}
where the symbol $\otimes $ stands for the convolution product.
\vspace{5 mm}

\noindent {\textbf{Confocal Imaging.}} 

The confocal image of the medium can be easily retrieved from the focused reflection matrix at each depth $z$ by considering the diagonal elements which verify $ \bm{\rho}_c=\bm{\rho}_{\textrm{in}} =\bm{\rho}_{\textrm{out}}$: 
\begin{equation}
\label{ImConf}
    \mathcal{I}(\bm{\rho}_c,z) = R(\bm{\rho}_c,\bm{\rho}_c,z)
\end{equation}
At each depth $z$, each line of the confocal image results from the convolution between sample reflectivity $\gamma$ and the
confocal PSF $H^2$ (Eqs.~\ref{RHH} and \ref{ImConf}):
\begin{equation}\label{ImConf2}
    \mathcal{I}(\bm{\rho}_c,z) = \int d\bm{\rho}{H}^2(\bm{\rho},\bm{\rho}_c,z) {\gamma}(\bm{\rho},z) 
\end{equation}
The confocal image is displayed in Fig.~\ref{figOrig}a but note that a time gain compensation has been priorly applied to get an homogeneous contrast over the whole depth range, as described below. 
\vspace{5 mm}

\noindent {\textbf{Depth gain compensation of the confocal image.}}

\noindent The raw confocal image actually exhibits a strong amplitude drop with depth (Supplementary Figure~9). This attenuation is due to the decay of energy experienced by seismic waves while they propagate. Without compensation, this attenuation strongly degrades the contrast of the confocal image at large depths.
The depth attenuation of the confocal signal can be caused  by several factors such as geometrical spreading, scattering and absorption (intrinsic or inelastic attenuation)~\cite{Shapiro1993,Aki1969}. In the present case, the geometrical spreading of waves is compensated, at least partially, by the focusing process performed both at input and output of the reflection matrix. The attenuation of the confocal image is thus mainly due to scattering and absorption. In a statistically homogeneous disordered medium, the mean intensity, $\langle |I(\bm{\rho},z)|^2\rangle $, shall scale as $\exp (-2 z/\ell_{\textrm{ext}})$. $\ell_{\textrm{ext}}$ is the extinction length that combines the scattering and absorption losses as follows: $\ell_{\textrm{ext}}^{-1}=\ell_{\textrm{s}}^{-1}+\ell_{\textrm{a}}^{-1}$, with $\ell_{\textrm{s}}$, the scattering mean free path and $\ell_{\textrm{a}}$, the absorption length. 

To retrieve such an exponential decay, the random-like fluctuations of the confocal image due to lateral reflectivity variations should be priorly smoothed out by averaging. The resulting mean confocal intensity, $\langle |I(\bm{\rho},z)|^2\rangle_{\bm{\rho}} $, is displayed in log-scale as a function of effective depth $z$ in Supplementary Figure 9. It highlights four depth ranges with distinct decay rates. For each depth range, the decrease of the mean confocal intensity is fitted by an exponential curve whose decay provides an estimation of $\ell_{\textrm{ext}}$ reported in Tab.~\ref{table1}.


The overall fitting curve, $\exp[-\beta (z)]$, displayed in  Supplementary Figure 9, can be used to normalize at each depth the confocal images shown in the manuscript, such that:
\begin{equation}
\label{norma}
\mathcal{I}_N(\bm{\rho},z)= \exp [\beta(z)/2] \mathcal{I}(\bm{\rho},z),
\end{equation}
with $\mathcal{I}_N(\bm{\rho},z)$, the normalized confocal image displayed in Fig.~\ref{figOrig}a.

\vspace{5 mm}

\noindent {\textbf{Reflection point spread function.}} 

Interestingly, the focused reflection matrix can provide a local assessment of the focusing quality. Lambert \text{et al.}~\cite{lambert_reflection_2020} showed that the amplitude distribution along each antidiagonal of $\mathbf{R}_{\bm{\rho}\bm{\rho}}(z)$ provides a key quantity that we will refer to as the reflection point-spread function (RPSF): 
\begin{equation}\label{Int_midpoint}
    RPSF(\Delta \bm{\rho},\bm{\rho}_c,z) = R(\bm{\rho}_c-\Delta\bm{\rho},\bm{\rho}_c+\Delta\bm{\rho})
\end{equation}
Along an antidiagonal of $\mathbf{R}_{\bm{\rho}\bm{\rho}}(z)$, all couple of points on a given antidiagonal share the same midpoint $\bm{\rho}_c = (\bm{\rho}\out + \bm{\rho}\inp)/2$ but with a varying relative position $\Delta \bm{\rho} = (\bm{\rho}\out - \bm{\rho}\inp)/2$. In the vicinity of an isolated scatterer at $(\bm{\rho}_s,z)$, the RSPF is a direct indicator of the local focusing quality (Supplementary Note \rev{7}):
\begin{equation}
\label{RPSF}
    RSPF(\Delta \bm{\rho},\bm{\rho}_s,z)=\exp \left (i \frac{k_0}{z} |\Delta \bm{\rho}|^2  \right) F \left (\frac{\Delta \bm{\rho}}{\lambda z}\right )  F \left (-\frac{\Delta \bm{\rho}}{\lambda z} \right)  .
\end{equation}
Therefore, the energy spreading in the vicinity of each scatterer position shall enable one to probe the spatial extension of the transmit PSF. As the scatterer positions are \textit{a priori} unknown, the RPSF is, in practice, probed by considering the antidiagonal whose common mid-point exhibits the maximum confocal signal.
\vspace{5 mm}

\noindent {\textbf{Compensation of aberrations.}} 

As highlighted above, the focused basis is the adequate framework for imaging and quantification of focusing quality. However, the reflection matrix shall be investigated into a dual basis to analyse and compensate for aberrations. 

\rev{The relevant basis for aberration correction depends on the nature of the medium heterogeneities. In a statistically random medium, the most adequate correction plane lies at $z/3$, a plane where the tilt and shift memory effect can be combined to maximize isoplanicity~\cite{Osnabrugge2017}. For a medium displaying wave velocity heterogeneities localized in depth, the best correction plane is the one conjugated with this aberrating layer. In the present case, we are in the latter situation with wave velocity heterogeneities arising close to the Earth surface.}

\rev{Therefore, the most adequate basis is here} a plane lying on the Earth surface, \textit{i.e} at the depth origin $z=0$ defined by the average elevation of the seismic stations $\mathbf{g}_i$. This plane is described by the coordinate vector $\mathbf{u}$. The broadband matrix can be projected in the Earth surface basis, first at its output, to yield the dual reflection matrix $\mathbf{R}_{\mathbf{u}\bm{\rho}}(z)=[R(\mathbf{u}_{\textrm{out}},\bm{\rho}_{\textrm{in}},z)]$. This projection can be performed by performing the following matrix product:
  \begin{equation}\label{Ruout}
     \mathbf{R}_{\mathbf{u}\bm{\rho}}(z)= \mathbf{G}_0^{\top}(z,f_0) \times  \mathbf{R}_{\bm{\rho}\bm{\rho}}(z)
 \end{equation}
 where the symbol $\top$ stands for matrix transpose. An angular de-scan of the input focusing points as sketched in Fig.~\ref{figCorrU}b can be performed by computing the Hadamard product between $\mathbf{R}_{\mathbf{u}\bm{\rho}}$ and its ideal counterpart $\mathbf{G}_{0}(z)$:
   \begin{equation}
   \label{Duout}
     \mathbf{D}_{\mathbf{u}\bm{\rho}}(z)= \mathbf{G}_0^{\dag}(z,f_0) \circ  \mathbf{R}_{\mathbf{u}\bm{\rho}}(z)
 \end{equation}
Each column of the resulting distortion matrix,  $\mathbf{D}_{\mathbf{u}\bm{\rho}}(z)=[D(\mathbf{u}_{\textrm{out}},\bm{\rho}_{\textrm{in}},z)]$, maps the phase-distortions \rev{with} respect to the ideal wave-front that would be obtained for a point-like source at $(\bm{\rho}_{\textrm{in}},z)$. 

An estimator $\mathbf{W}_\mathbf{u}$ of the aberration transmittance is then extracted through an iterative phase reversal (IPR) process applied to the correlation matrix $\mathbf{C}_{\mathbf{u}\mathbf{u}}=\mathbf{D}_{\mathbf{u}\bm{\rho}} \times \mathbf{D}_{\mathbf{u}\bm{\rho}}^{\dag}$ (see below). The phase conjugate of the estimator $\mathbf{W}_\mathbf{u}$ is then used as a focusing law to compensate (partially) for wave distortions. An updated focused reflection matrix is obtained through the following relation: 
\begin{equation}
    \mathbf{R}_{\bm{\rho} \bm{\rho}}(z)=\mathbf{G}_0^{\dag}(z,f_0) \times \left [ \mathbf{G}_0(z,f_0) \circ \mathbf{W}^*(z) \circ  \mathbf{D}_{\mathbf{u}\mathbf{\rho}}(z) \right ] 
\end{equation}

The whole process is then iterated to improve the estimation of the aberration transmittance by alternating aberration correction at input and output~\cite{lambert_distortion_2020}. In practice, two iterations of the aberration correction process were enough to converge in the present case.

At the end of the process, a novel confocal image is obtained by considering the diagonal elements of the updated focused reflection matrix [Figs.~\ref{figCorrU}e and f]. The fine compensation of wave distortions is highlighted by the RSPF (Eq.~\ref{Int_midpoint}) deduced from the updated focused $\mathbf{R}-$matrix (Fig.~\ref{figCorrU}g). As expected theoretically, compensation of aberrations in the geophone basis enables the recovery of a resolution only limited by the geophone aperture. As shown by Fig.~\ref{figCorrU}e, this is nevertheless not sufficient to have a contrasted image of the volcano in depth.

\vspace{5 mm}

\noindent {\textbf{Compensation of diffraction.}}

To go beyond and beat diffraction, the parabolic phase law exhibited by the RPSF (Fig.~\ref{figCorrU}h) can be exploited (Supplementary Note \rev{7}). To that aim, the $\mathbf{R}-$matrix shall be investigated between the focused basis and the $\mathbf{k}-$space. The focused reflection matrix is thus projected in the plane wave basis, first at output, such that:
\begin{equation}\label{Rkout}
     \mathbf{R}_{\mathbf{k}\bm{\rho}}(z)= \mathbf{T_0} \times  \mathbf{R}_{\bm{\rho}\bm{\rho}}(z),
\end{equation}
with $\mathbf{T}_0$, the Fourier transform operator
 \begin{equation}\label{T0kr}
    T_0(\mathbf{k}_{||},\bm{\rho}) = \exp\left( - i \mathbf{k}_{||}.\bm{\rho} \right).
\end{equation}
Then, a new distortion matrix can be built in the plane wave basis by comparing the each reflected wave-field, $R(\mathbf{k}_{\textrm{out}},\bm{\rho}_{\textrm{in}})$, in the $\mathbf{k}$ space with its reference counterpart, $T_0(\mathbf{k}_{\textrm{out}},\bm{\rho}_{\textrm{in}})$, that would be obtained for a point-like guide star at $(\bm{\rho}_{\textrm{in}},z)$:
    \begin{equation}\label{Dkout}
     \mathbf{D}_{\mathbf{k}\bm{\rho}}(z)=   \mathbf{R}_{\mathbf{k}\bm{\rho}}(z) \circ \mathbf{T}_0^*(z)
\end{equation}
From this matrix $\mathbf{D}_{\mathbf{k}\bm{\rho}}$, a diffraction transmittance $\mathbf{W}_{\mathbf{k}}$ (Fig.~\ref{figCorrK}c) can be extracted by applying the IPR process to the correlation matrix $\mathbf{C}_{\mathbf{k}\mathbf{k}}=\mathbf{D}_{\mathbf{k}\bm{\rho}} \times \mathbf{D}_{\mathbf{k}\bm{\rho}}^{\dag}$ (see below).

The phase conjugate of $\mathbf{W}_\mathbf{k}$ is then used as a focusing law to compensate for wave diffraction (Supplementary Note \rev{9}). An updated focused reflection matrix is obtained through the following relation: 
\begin{equation}
    \mathbf{R}_{\bm{\rho} \bm{\rho}}(z)=\mathbf{T}_0^{\dag} \times \left [ \mathbf{T}_0 \circ \mathbf{W}_{\mathbf{k}}^*(z) \circ  \mathbf{D}_{\mathbf{k}\mathbf{\rho}}(z) \right ] 
\end{equation}

The whole process is then repeated at input to compensate for the diffraction phenomena undergone by the down-going wave-fields. At the end of the process, a novel confocal image is obtained by considering the diagonal elements of the updated focused reflection matrix [Figs.~\ref{figCorrK}d and e]. The beating of diffraction is highlighted by the RSPF (Eq.~\ref{Int_midpoint}) deduced from the updated focused $\mathbf{R}-$matrix (Fig.~\ref{figCorrK}f). The resolution reaches the ultimate diffraction limit ($\sim \lambda/2$) and is no longer limited by the geophone network aperture (white circle in Fig.~\ref{figCorrK}f).

\vspace{5 mm}

\noindent {\textbf{Iterative phase reversal}} 

The IPR algorithm is a computational process that provides an estimator of the aberration and/or diffraction transmittance at each depth $z$ extracted from the correlation matrices computed at the Earth surface ($\mathbf{C}_{\mathbf{u}\mathbf{u}}$) and in the plane wave basis ($\mathbf{C}_{\mathbf{k}\mathbf{k}}$), respectively. Mathematically, the algorithm is based on the following recursive relation:
\begin{equation}
\label{IPsR}
    \mathbf{W}_{\mathbf{x}}^{(n+1)}=\exp \left ( i \mbox{arg} \left \lbrace \mathbf{C_{xx}} \times \mathbf{W}_{\mathbf{x}}^{(n)} \right \rbrace \right )
\end{equation}
with $\mathbf{x}=\mathbf{u}$ or $\mathbf{k}$, the coordinate vector in the correction basis. $\mathbf{W}_{\mathbf{x}}^{(0)}=[1 \cdots 1]^T$ is chosen arbitrarily as a unit wave-front. The resulting wave-front at the end of the IPR process,  $\mathbf{W}_{\mathbf{x}}=\lim\limits_{n\rightarrow \infty}\mathbf{W}_{\mathbf{x}}^{(n)}$, provides an estimator of the aberration transmittance in the Earth basis (Supplementary Note \rev{8}) and of the diffraction transmittance in the plane wave basis (Supplementary Note \rev{9}).
\vspace{5 mm}

\noindent\textbf{Data availability.} Seismic data used in this manuscript has been deposited at the \alex{\href{https://doi.org/10.18715/GUADELOUPE.OVSG}{Data collection of the seismological and volcanological observatory of Guadeloupe}~\cite{IDPDGDP2021,Burtin2017}}. The seismic noise correlation data generated in this study are available at Zenodo~\cite{Giraudat2023} (\href{https://zenodo.org/record/10066910}{https://zenodo.org/record/10066910}).
\vspace{5 mm}

\noindent\textbf{Code availability.}
The codes used to post-process the noise correlation data are available at Zenodo~\cite{Giraudat2023} (\href{https://zenodo.org/record/10066910}{https://zenodo.org/record/10066910}).

\bibliography{scibib.bib}

\clearpage

\noindent\textbf{Acknowledgments.}
The authors wish to thank R. Touma, M. Campillo and A. Derode for initial discussions on the project{; the colleagues at the OVSG-IPGP~\cite{IDPDGDP2021} for field assistance in installing and running the node network, SISMOB network of RESIF (now Epos-France); the Parc National de Guadeloupe for permission to install the node in the field.} The authors are grateful for the funding provided by the European Research Council (ERC) under the European Union's Horizon 2020 research and innovation program (grant agreement no. 819261, REMINISCENCE project). This work has also been supported by the AO-IPGP 2017 project ``Dense seismic monitoring of the hydrothermal system of La Soufrière de Guadeloupe'', the AO-TelluS-INSU 2017 action ALEAS (coord.: A. Burtin), the project ``Vers la Plateforme R\'{e}gionale de Surveillance Tellurique
du futur'' - (PREST) co-funded by INTERREG Caraïbes V for the European Regional Development Fund, and the European Union's Horizon 2020 research and innovation programme (grant agreement, no. 731070, EUROVOLC project). The authors also thank IPGP for general funding to the Observatoires Volcanologiques et Sismologiques (OVS), the INSU-CNRS for funding provided by the Service National d'Observation en Volcanologie (SNOV), and the \rev{Ministère de la transition \'{e}cologique et de la coh\'{e}sion des territoires (MTECT)} as well as the IdEx project ``Universit\'{e} Paris Cit\'{e}'' (ANR-18-IDEX-0001)  for financial support. 
\vspace{5 mm}

\noindent\textbf{Author Contributions.}
A.A. and A.B. initiated the project. A.B. designed and conducted the collection of seismic data. A.B. performed the cross-correlation of seismic data. E.G., A.L.B, and A.A. developed the post-processing tools.  E.G. and A.A. performed the theoretical analysis. E.G. prepared the figures. A.B. and J.-C. K. provided the geophysical and {volcanological} interpretation for the obtained image. E.G., {J.-C. K.} and A.A. prepared the manuscript. E.G., A.B., A.L.B., J.-C. K., M.F., and A.A. discussed the results and contributed to finalizing the manuscript.\\

\noindent\textbf{Competing interests.}
The authors declare no competing interests.
 
 \clearpage

\noindent\textbf{Tables}
\vspace{5mm}

\begin{table}[!ht]
   \center
   \small
\begin{tabular}{|c|c|}
    \hline
    {Depth range} & {Extinction length}  \\
  \hline \hline
   0 - 0.5 km & 1665 m \\
   \hline
  0.5 - 1.9 km & 310 m \\
  \hline
  1.9 - 3.8 km  & 765 m \\
  \hline
  3.8 - 10.5 km & 3070 m \\
   \hline
\end{tabular}
\caption{\textbf{Extinction length}. Estimation of the extinction length $\ell_{\textrm{ext}}$ from the depth decay of the confocal intensity displayed in Supplementary Figure 9.}
\label{table1}
\end{table}

\clearpage

\vspace{\baselineskip}

\clearpage 

\clearpage

\renewcommand{\thetable}{S\arabic{table}}
\renewcommand{\thefigure}{S\arabic{figure}}
\renewcommand{\theequation}{S\arabic{equation}}
\renewcommand{\thesection}{S\arabic{section}}

\setcounter{equation}{0}
\setcounter{figure}{0}
\setcounter{section}{0}

\begin{center}
\huge{\bf{Supplementary Information}}
\end{center}
\normalsize
\vspace{5 mm}

\textbf{This document provides further information on: \rev{ (\textit{i}) the convergence of the seismic noise correlation functions; (\textit{ii}) the main seismic noise sources in La Soufrière; (\textit{iii}) the asymmetry of the noise correlation functions; (\textit{iv}) the filtering of spurious arrivals with the redatuming process; }(\textit{v}) the depth evolution of the scattering intensity; (\textit{vi}) the transmit point spread function; (\textit{vii}) the reflection point spread function; (\textit{viii}) the iterative phase reversal algorithm for compensation of wave distortions from the Earth surface basis; (\textit{ix}) the iterative phase reversal driven from $\mathbf{k}$-space and its comparison with a singular value decomposition approach; (\textit{x}) the spatial resolution of the final image; (\textit{xi}) the depth evolution of the maximum confocal signal}

\section{Noise correlation convergence}
\label{sec1}
\rev {In this Supplementary Note, we investigate the convergence of the NCF towards their final value. Let $n_i(\tau)$ be the 1-bit and spectrally-whitened noise recorded by the $i^{\textrm{th}}$ geophone, $\tau$ denoting the absolute time. The NCF, $\Gamma_{ij}(t,T)$, for each couple of geophone such that $i<j$ can be computed over an integration time $T$:
\begin{equation}
\label{corrGamma}
\Gamma(\mathbf{g}_i,\mathbf{g}_j,t,T)=\frac{1}{T}\int n_i(\tau) n_j(\tau+t) d\tau. 
\end{equation}
The convergence of each NCF towards its finite value, \begin{equation}
R(\mathbf{g}_i,\mathbf{g}_j,t)=\lim\limits_{T \rightarrow T_{\textrm{max}}} \Gamma(\mathbf{g}_i,\mathbf{g}_j,t,T), 
\end{equation}
can be monitored by computing the following scalar product:
\begin{equation}
\label{Psca}
P(\mathbf{g}_i,\mathbf{g}_j,T)=\int_{-t_{\textrm{max}}}^{t_{\textrm{max}}} dt R(\mathbf{g}_i,\mathbf{g}_j,t) \Gamma(\mathbf{g}_i,\mathbf{g}_j,t,T). 
\end{equation}
The corresponding scalar product is displayed as a function of $T$ for each geophone couple in Supplementary Figure~\ref{convergence2}. Its averaged value $\bar{P}(T)=\langle  P(\mathbf{g}_i,\mathbf{g}_j,T)\rangle_{\lbrace \mathbf{g}_i,\mathbf{g}_j \rbrace }$ over each geophone couple shows a constant increasing rate with a characteristic convergence time of 2.5 days to reach a 90\% threshold. }
\clearpage 

\begin{figure*}[h]   
\centerline{\includegraphics[width = 12 cm]{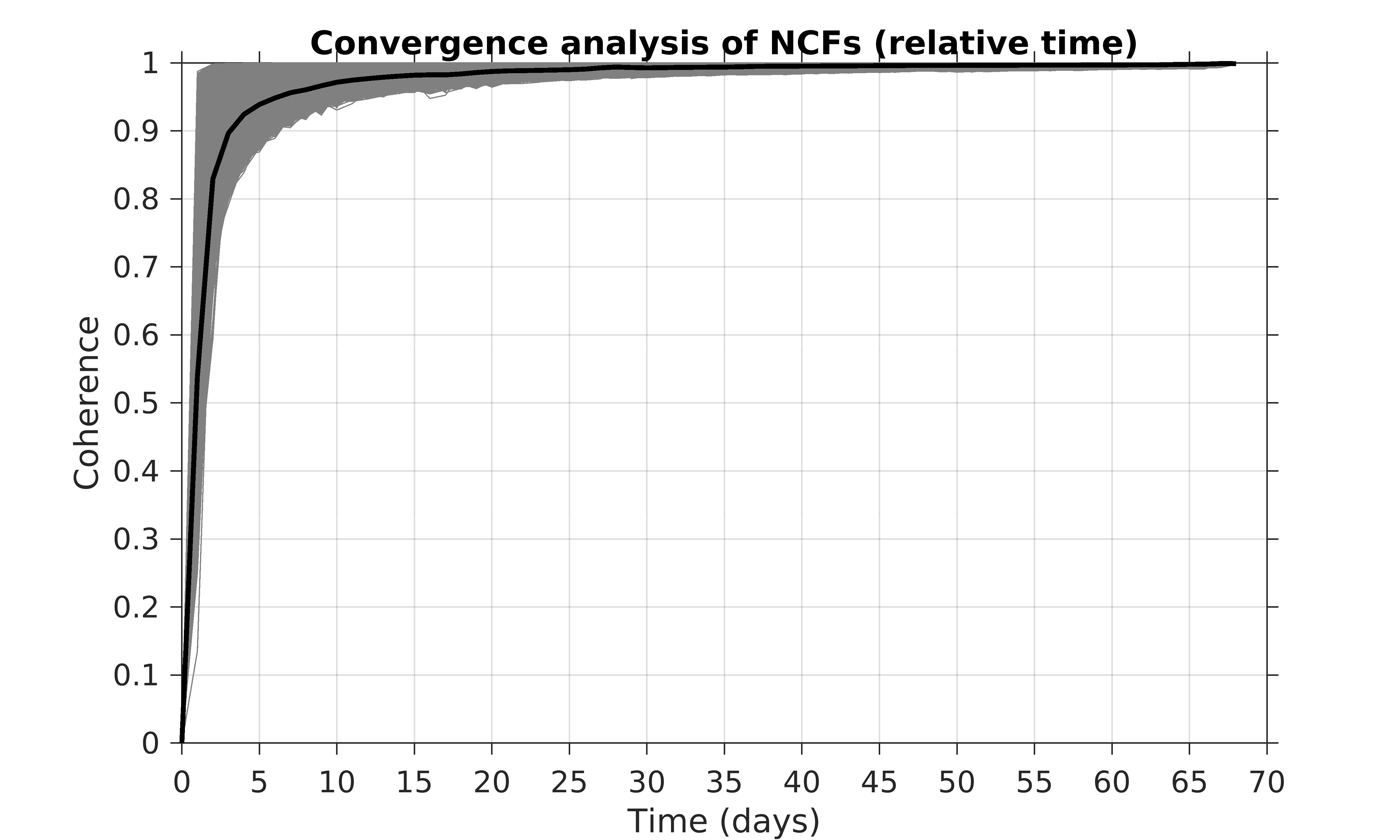}}
\caption{\rev{\textbf{Convergence of the noise correlation functions}. The scalar product (Eq.~\ref{Psca}) between each NCF computed over integration time $T$ and its final value (2 months-correlation) is plotted as a function of $T$ (gray lines). The mean scalar product is plotted as a black line.}}
\label{convergence2}
\end{figure*}

\begin{figure*}[h]   
\centerline{\includegraphics[width=12 cm]{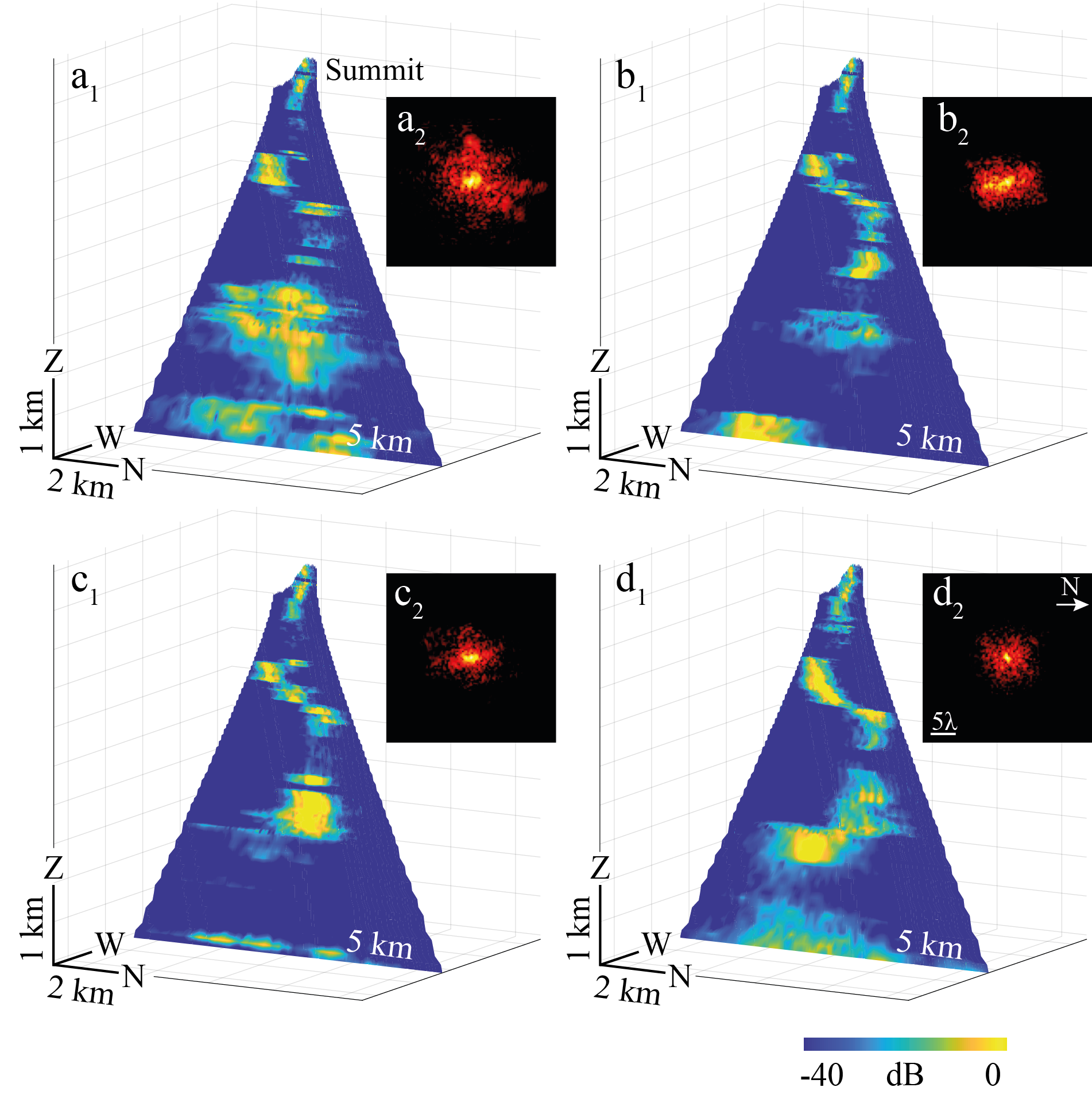}}
\caption{\rev{\textbf{Matrix image for several integration times}. Result of the matrix imaging process using NCFs computed over different integration times: \textbf{a}, $T=2$ days; \textbf{b}, $T=5$ days; \textbf{c}, $T=10$ days; \textbf{d}, $T=2$ months. For each panel, we show: ($1$) the vertical slice of the 3D matrix image along the South-North direction with the same view as in Fig.~5d of the accompanying paper but the image is restricted to the first $5$~km below the summit for a clearer visualization; ($2$) an example of corrected RPSF at $z=3500$ m.}}
\label{convergence_image}
\end{figure*}
\rev{Nevertheless, one has to be cautious with this value since the scalar product $P$ is considered over the whole time-of-flight range. The convergence of the NCF is \textit{a priori} slower for long times-of-flight. The amplitude of reflected waves actually decreases with the time-of-flight $t$, or equivalently, the corresponding ballistic depth $z$ (Supplementary Figure~\ref{normZ}). This effect is highlighted by Supplementary Figure~\ref{convergence_image} that displays the final matrix image of the upper part of the volcano using NCFs computed over different correlation times: $T=2$ days (Supplementary Fig.~\ref{convergence_image}a), $T=5$ days (Supplementary  Fig.~\ref{convergence_image}b), $T=10$ days (Supplementary Fig.~\ref{convergence_image}c), $T=2$ months (Supplementary Fig.~\ref{convergence_image}d).
While the upper part of the volcano (1000 m), in particular the top part of the conduit, is retrieved after a few days of correlations, the magma storage zone beyond 5 km is only recovered by computing the correlation over two months. It means that, for a given integration time, the signal-to-noise ratio in the final image will decrease with the inspected depth. In perspective of dynamic seismic imaging, the absolute resolution time will therefore depend on this depth as well. While fluid motion at shallow depth can be monitored in a day-to-day framework, the control of the magma storage zone could only be performed at a much longer time scale, of the order of one month. }   

\rev{Supplementary Fig.~\ref{convergence_image} also shows the impact of the NCFs integration time on the RPSF. While the RPSF shows an important incoherent background for $T=2$ days (Supplementary Fig.~\ref{convergence_image}a$_2$), this background gradually decreases with the integration time $T$, as shown by Supplementary Figs.~\ref{convergence_image}b$_2$, c$_2$ and d$_2$. The quality of the NCFs has therefore an impact on the RPSF background. The latter observable can thus be an adequate tool to monitor the convergence of the NCFs.}
\clearpage

\section{Noise sources}
\label{sec2}
\rev{The convergence of the NCF towards a finite value (Supplementary Figure~\ref{convergence2}) does not mean that they do converge towards the Green's functions. Indeed, such a convergence is only fulfilled under energy equipartition, \textit{i.e} an homogeneous distribution of seismic noise energy in phase space~\cite{Weaver1990}. This condition can be reached with a randomly distributed noise source distribution but is quite restrictive. Fortunately, multiple scattering can help approaching energy equipartition~\cite{Hennino2001}. In the following, we investigate the different seismic noise sources and their characteristics in La Soufrière.}

\begin{figure*}[h]   
\centerline{\includegraphics[width = 12 cm]{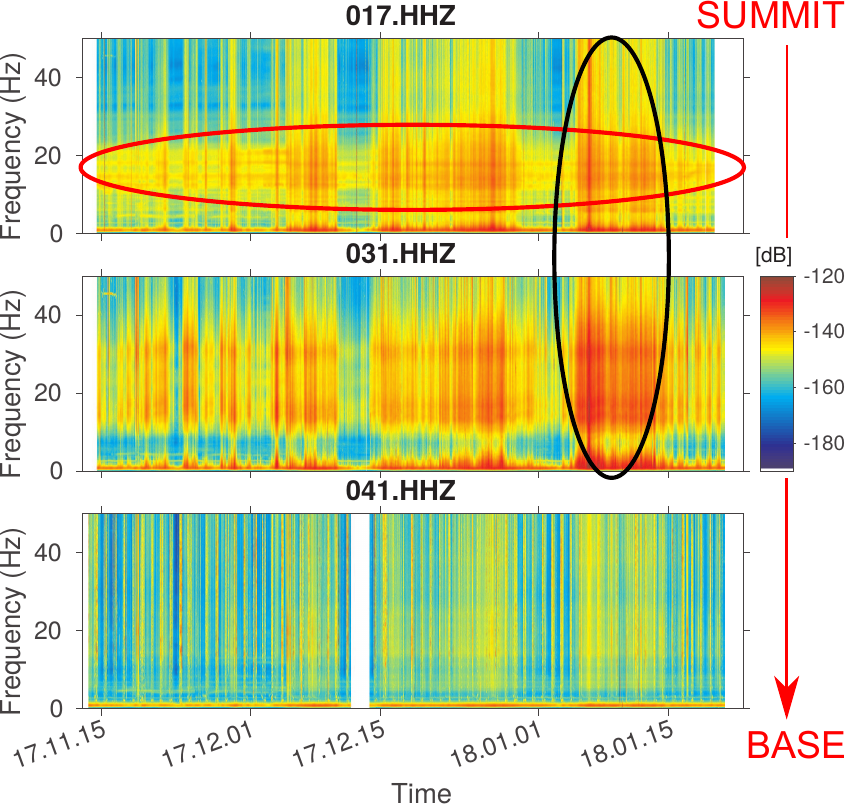}}
\caption{\rev{\textbf{Seismic noise recorded by three nodes during the two months-measurement}. Spectrogram versus time of the seismic noise recorded by: \textbf{a}, node $17$ placed at the summit of La Soufrière; \textbf{b}, node $31$ lying on the side of the volcano; \textbf{c}, node $41$ located at its base. }}
\label{spectro}
\end{figure*}

\rev{The main noise sources in La Soufrière can be revealed by investigating the spectrogram of the seismic noise measured by three nodes located at three locations of the volcano displayed in Supplementary Fig.~\ref{spectro}. Three different noise sources can be discriminated in the 10-20 Hz-frequency bandwidth:
\begin{itemize}
\item A continuous noise source associated with fumaroles located close to the summit of the volcano:  Localized at the north of the geophone array, the fumaroles are associated with a peculiar frequency content that manifests as horizontal lines surrounding by a red ellipse in Supplementary Fig.~\ref{spectro}a.
\item A daylight broadband noise source generated by anthropic activity: Generated by the Basse-Terre city, located at 10 km South West of the geophone network, this noise source is associated with the vertical strips exhibited by the spectrogram of the station at the base of the volcano (see Supplementary Fig.~\ref{spectro}c).
\item A more diffuse source of noise generated by the wind that manifests as vertical bands of random duration on the spectrogram, each band corresponding to specific weather events (see one example surrounded by a black ellipse in Supplementary Fig.~\ref{spectro}).
\end{itemize}}

\section{Noise correlation functions}
\label{sec3}

\begin{figure*}[h]   
\centerline{\includegraphics[width = 16 cm]{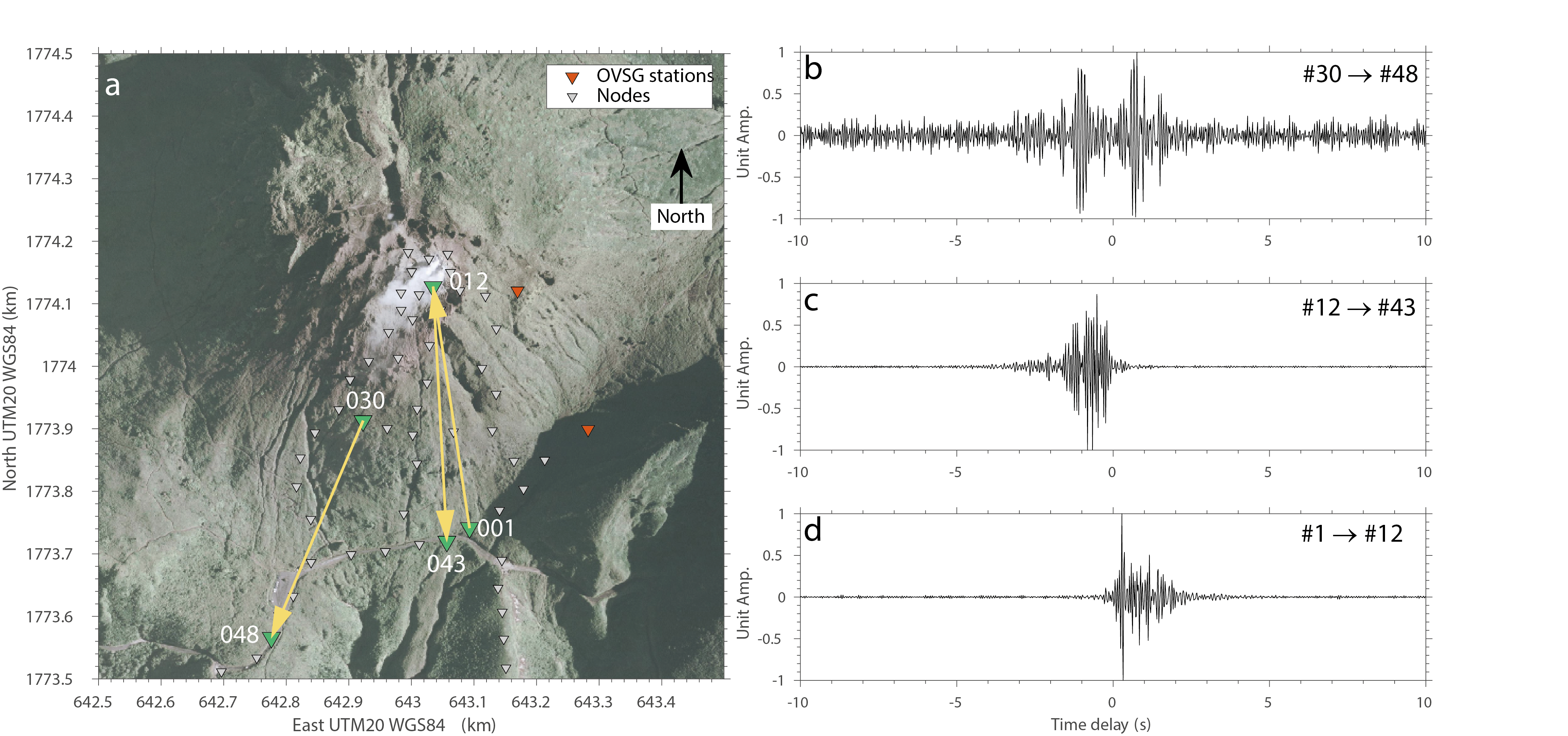}}
\caption{\rev{\textbf{Impact of fumaroles on the NCFs}. \textbf{a}, Location of nodes \#1, \#12, \#30, \#43 and \#48 whose cross-correlations functions are displayed in panels b-d. Map data: Google, CNES, Airbus, 2023. \textbf{b}, NCF between nodes \#30 and \#48. \textbf{c}, NCF between nodes \#1 and \#12. \textbf{d}, NCF between nodes \#12 and \#43.}}
\label{fumarole}
\end{figure*}

\rev{Let us now investigate the impact of this heterogeneous noise source distribution on the NCF. While, for some couples of geophones, energy equipartition seems to be fulfilled with a nearly symmetric NCF (\#30 $\rightarrow$ \#48, Supplementary Fig.~\ref{fumarole}b), this is far from being the case in general. Supplementary Figs.~\ref{fumarole}c and d show the NCFs between two couples of stations oriented along the South-North direction (\#1$\rightarrow$\#12, Supplementary Fig.~\ref{fumarole}b) and along the North-South directions (\#12$\rightarrow$\#43, Supplementary Fig.~\ref{fumarole}c), respectively. While only an anti-causal component is observed in the former case, a causal wave-field is obtained in the latter case. This strong asymmetry between the two NCFs is a manifestation of the fumaroles localization right at the north of the array network~\cite{Schippkus2022}. When the two stations are aligned with an isolated noise source, a fully causal or anti-causal correlation function is obtained as sketched in Supplementary Figs.~\ref{convention}a,b). More generally, when not aligned with the two stations, a localized noise source leads to spurious anti-causal or causal arrivals emerging before the ballistic time~\cite{Schippkus2022}.}

\begin{figure*}[h]   
\centerline{\includegraphics[width = 10 cm]{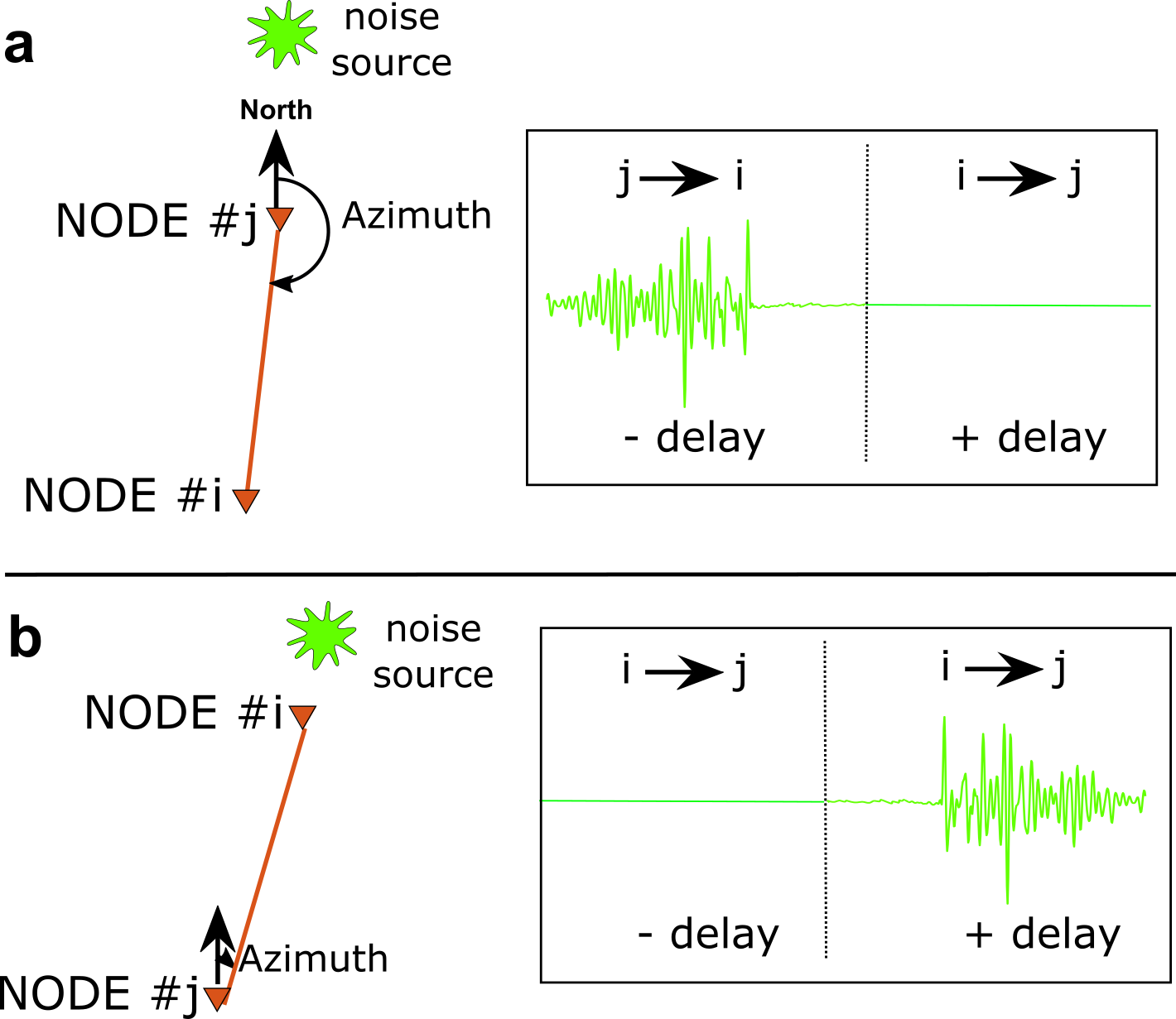}}
\caption{\rev{\textbf{Correlation function in presence of a localized noise source}. \textbf{a}, For a node couple $(i,j)$ oriented along the South-North direction, the ballistic wave induced by a noise source at North will emerge along the anti-causal component of the NCF. \textbf{b}, For a node couple $(i,j)$ oriented along the South-West direction, the ballistic wave induced by a noise source at North-East (green) will emerge along the causal component of the NCF.}}
\label{convention}
\end{figure*}

\begin{figure*}[h]   
\centerline{\includegraphics[width = 16 cm]{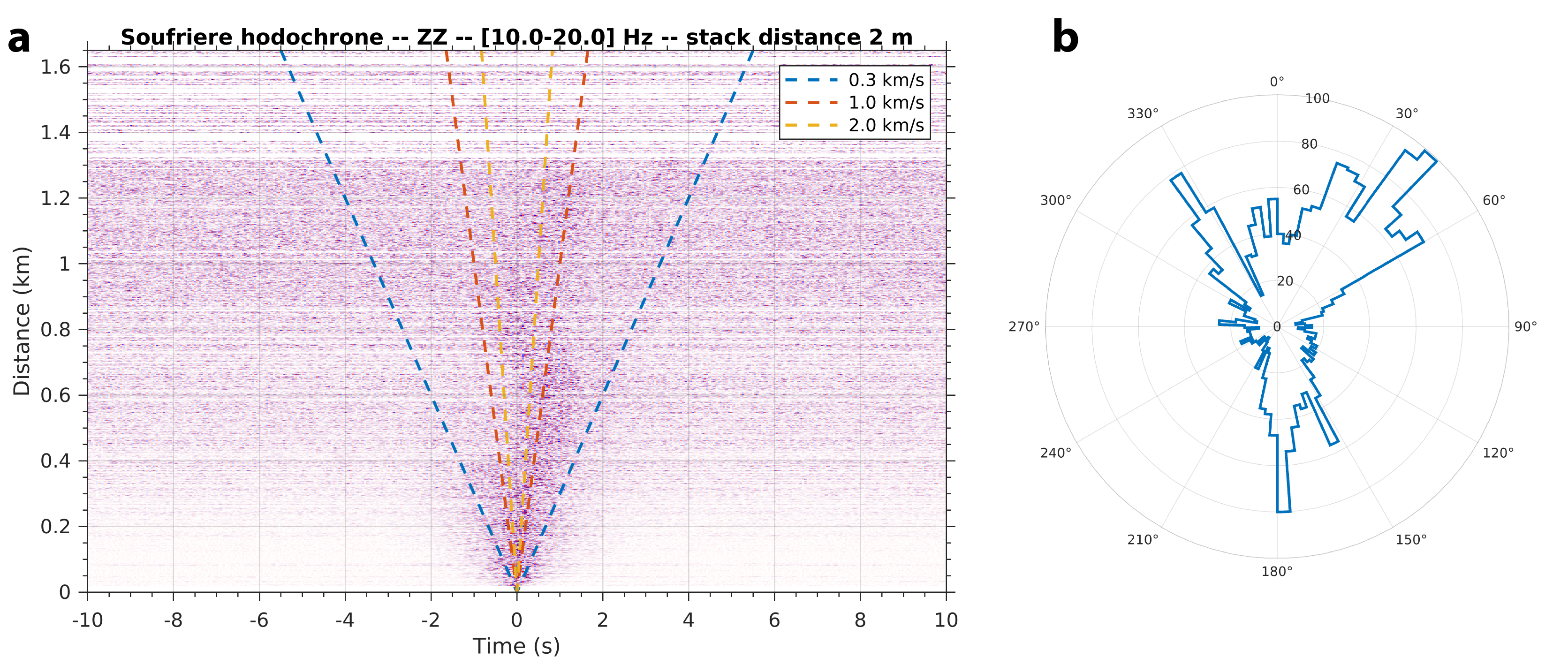}}
\caption{\rev{\textbf{Hodochrone representation of correlation functions}. \textbf{a}, The correlation functions are plotted as a function of the time lapse and lateral distance between geophones. \textbf{b}, Azimuth angle distribution of source-receiver couples in the NCFs using the azimuth angle convention shown in Supplementary Fig.~\ref{convention}.}}
\label{hodochrone}
\end{figure*}

\rev{Beyond these specific examples, a predominance of the causal component is observed when stacking the whole set of NCFs (Supplementary Fig.~\ref{hodochrone}a). To understand this feature, it should be first reminded that each NCF between two geophones $i$ and $j$ is computed for $i<j$ (Eq.~\ref{corrGamma}). For the causal component, the geophone associated with the smaller index $i$ is therefore the virtual source and the geophone associated with the larger index $j$ is the virtual receiver. For the anti-causal component, the roles of nodes $i$ and $j$ are exchanged. This convention is sketched in Supplementary Figs.~\ref{convention}a and b. The azimuth distribution of the virtual source-receiver couples for the geophone network is also shown in Supplementary Fig.~\ref{hodochrone}b. Its elongated shape translates the North-South expansion of the array. The predominance of the upper quadrant implies that virtual sources $i$ are usually at north of node $j$. This convention and the presence of fumaroles at the north of the geophone network therefore account for the predominance of the causal component displayed by the NCF in Supplementary Fig.~\ref{hodochrone}a.}

\rev{To avoid the spurious arrivals due to fumaroles, only the anti-causal component of the NCF has been considered for imaging the volcano. Yet the stacked anti-causal component shows a random wave-field without a clear signature of ballistic arrivals or reflected wave packets. This is due to the particular complex topography of a volcano, the heterogeneities of its subsoil and the absence of coherent reflectors. This random wave-field explains the blurred confocal image displayed in Fig.~2 of the accompanying paper but also highlights the feat of matrix imaging in revealing the inner structure of the volcano from such a complex wave-field. }

\begin{figure*}[h]   
\centerline{\includegraphics[width = 16 cm]{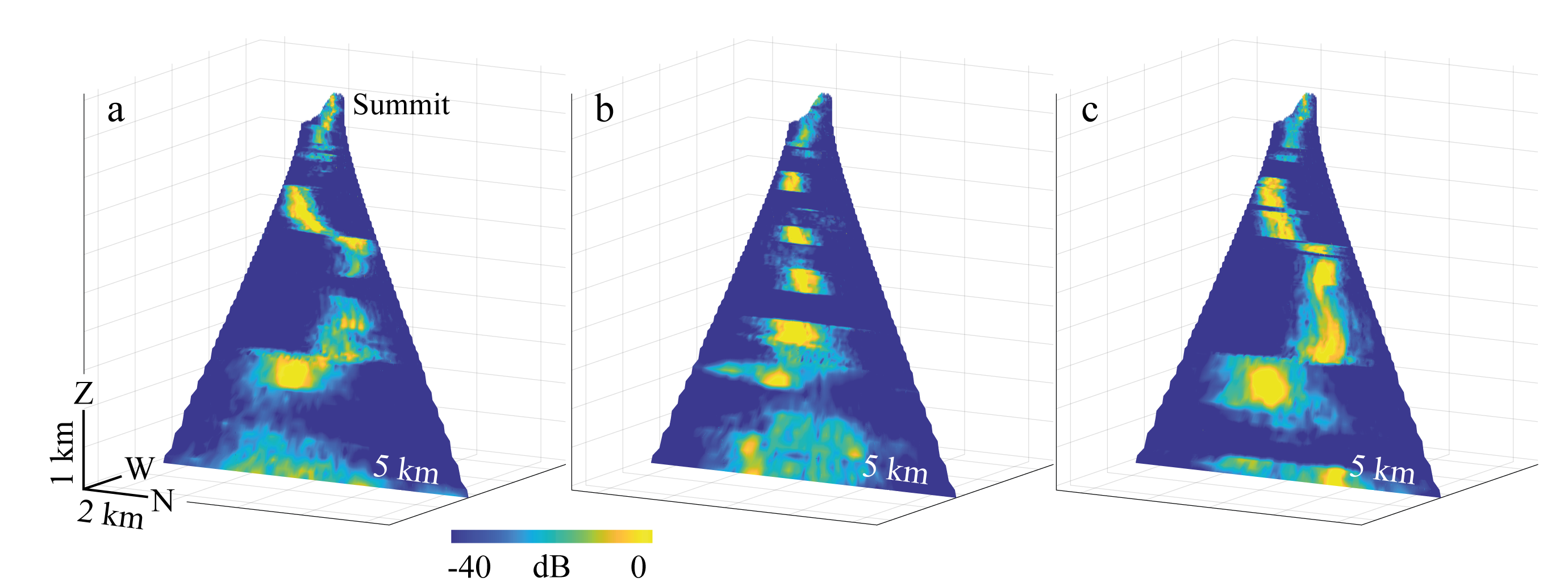}}
\caption{\rev{\textbf{Impact of NCFs on the final image of the volcano}. \textbf{a}-\textbf{b}, Matrix image of the volcano using the anti-causal and causal parts of the NCFs, respectively. \textbf{c}, Matrix image of the volcano using symmetrized NCFs. The view is identical to Figs.~2a, 4e and 5d of the accompanying paper, \rev{but the image is here restricted to the first $5$~km below the summit for a clearer visualization.} }}
\label{causal}
\end{figure*}

\rev{Supplementary Figs.~\ref{causal}a and b show the matrix images obtained by considering the anti-causal and causal components of the NCFs, respectively. While the anti-causal wave-field provides a matrix image displaying a strong continuity in depth, the causal wave-field gives rise to a more erratic behavior. Our interpretation of this difference is that the causal part of the Green's function is mainly associated with the fumaroles. This noise source is extremely localized and leads to an imperfect convergence towards the Green's function. On the contrary, the anti-causal component mainly originates from: (\textit{i}) an anthropic noise source distribution which is distant and thus more diffuse; (\textit{ii}) a randomly distributed noise distribution generated by the wind. Moreover, the scattered component of the NCFs exploits multiple scattering paths that allows a better energy equipartition and should provide a better convergence towards the anti-causal part of the Green's functions between geophones. }

\rev{The anti-causal component thus leads to a more reliable image of the volcano reflectivity, with a satisfying continuity of the image in depth. Note that a forced symmetrization of the NCFs by summing the NCFs and their time-reversed countepart does not lead to a better result (Supplementary Fig.~\ref{causal}c). The comparison between the different images displayed in Supplementary Fig.~\ref{causal} justifies \textit{a posteriori} our choice of only considering the anti-causal component of the NCFs for imaging the volcano. }

\section{Filtering the spurious arrivals with the redatuming process}
\label{sec4}

\begin{figure*}[h]   
\centerline{\includegraphics[width = 16 cm]{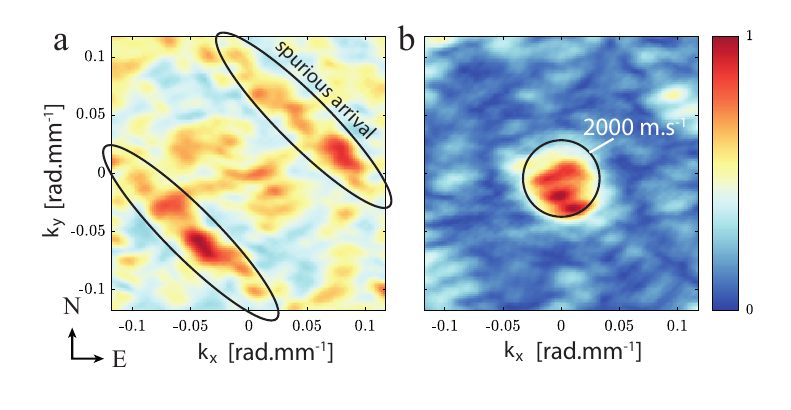}}
\caption{\rev{\textbf{Filtering of spurious arrivals by the redatuming process}. Apparent slowness of seismic echoes contained in the response matrix. \textbf{a}, Plane wave decomposition of the reflection matrix $\overline{\mathbf{R}}_{\mathbf{gg}}$ at output ($f=15$ Hz, Eq.~\ref{Rsk_0}) averaged over the set of geophones at input. Main echoes associated with spurious surface waves are surrounded by black ellipses.  \textbf{b} Plane wave decomposition of the filtered response matrix $\overline{\mathbf{R'}}_{\mathbf{gg}}$ at output ($f=15$ Hz, z=3600 m, Eq.~\ref{K_prim}) averaged over the set of geophones at input. The black circle correspond to an apparent velocity of $2000$ m/s.}}
\label{spurious}
\end{figure*}

\rev{Even though energy equipartition has been favored by considering only the anti-causal component of the NCFs, some spurious arrivals can still pollute the synthesized reflection matrix. A way to highlight their presence is to investigate the response matrix $\mathbf{R_{gg}}$ in the frequency domain. To this aim, a temporal Fourier transform is applied to $\mathbf{R_{gg}}(t)$. For each frequency $f$ in the bandwidth of interest (10-20 Hz), a monochromatic matrix $\overline{\mathbf{R}}_\mathbf{gg}(f)$ is obtained. The different wave components can then be discriminated by a plane wave decomposition of the output wave-fields, such that
\begin{equation}
\label{Rsk_0}
{\overline{\mathbf{R}}_\mathbf{kg}(f)= \mathbf{P}_{0}^T \times \overline{\mathbf{R}}_\mathbf{gg} (f) ,}
\end{equation}
{where the symbol $\times$ stands for the standard matrix product}. $\mathbf{P}_0=[P_0({\mathbf{g},\mathbf{k}})]$ is the Fourier transform operator that connects each geophone's position $\mathbf{g}$ to the transverse wave vector $\mathbf{k}=(k_x,k_y)$ of each angular component of the wave-field:
 \begin{equation}
{P}_0(\mathbf{g},\mathbf{k}) = \exp{\left(i {\mathbf{k}}\cdot \mathbf{g}_{||} \right), }
\label{Gsk}
\end{equation}
where the symbol $\cdot $ denotes the scalar product and $\mathbf{g}_{||} $ the transverse component of the position vector $\mathbf{g}$. Supplementary Fig.~\ref{spurious}a shows the result of this plane wave decomposition by displaying the mean angular distribution of the output wave-field at $f=15$ Hz. In that representation, surface Rayleigh waves should emerge along a circle whose radius would correspond to their slowness. As with the hodochrone representation of NCFs (Supplementary Fig.~\ref{hodochrone}a), we do not see any clear feature of Rayleigh waves, probably because of the complex topography of La Soufrière. Instead, we see the clear signature of spurious arrivals (black ellipses) probably caused by a localized anthropogenic noise coming from Basse Terre city located at South West of the geophone network. Besides those spurious arrivals, Supplementary Fig.~\ref{spurious}a seems to show bulk wave components arising at small spatial frequencies and corresponding to nearly vertical echoes. Nevertheless, it is difficult at this stage to be more affirmative since an imperfect convergence of NCFs could also lead to such a random wave-field. }  

\rev{Now we show how the redatuming process acts as a low-pass filter in the spatial frequency domain, thereby removing the spurious arrivals and selecting the reflected bulk waves. To illustrate this phenomenon, one can back-project the focused reflection matrix in the geophone basis,
\begin{equation}
\label{K_prim}
{\overline{\mathbf{R}}'_{\mathbf{gg}}(f)=\mathbf{G}^{\top}_{0}(z,f) \times \overline{\mathbf{R}}_{\bm{\rho \rho}}(z,f) \times \mathbf{G}_{0}(z,f)},
\end{equation}
From $\overline{\mathbf{R}}'_{\mathbf{gg}}(f)$, one can investigate the angular decomposition of the output wave-fields as previously done for the original response matrix $\overline{\mathbf{R}}_{\mathbf{gg}}$ (Eq.~\ref{Rsk_0}). The result is displayed in Supplementary Fig.~\ref{spurious}b. The comparison with its original counterpart in Supplementary Fig.~\ref{spurious}a highlights the low pass-filter operated by redatuming~\cite{Touma2023}: The spurious arrivals are discarded and only the P-waves associated with spatial frequencies $k<2\pi f/c_0$ are kept.  Several wave packets can be discerned and are associated with peculiar single scattering events at the focal depth ($z=3600$ m in the present case). Matrix imaging should be then applied to build a reflectivity image from these bulk wave echoes.}

\section{Scattering intensity vs. Depth}

\begin{figure*}[h]   
\centerline{\includegraphics[width = 16 cm]{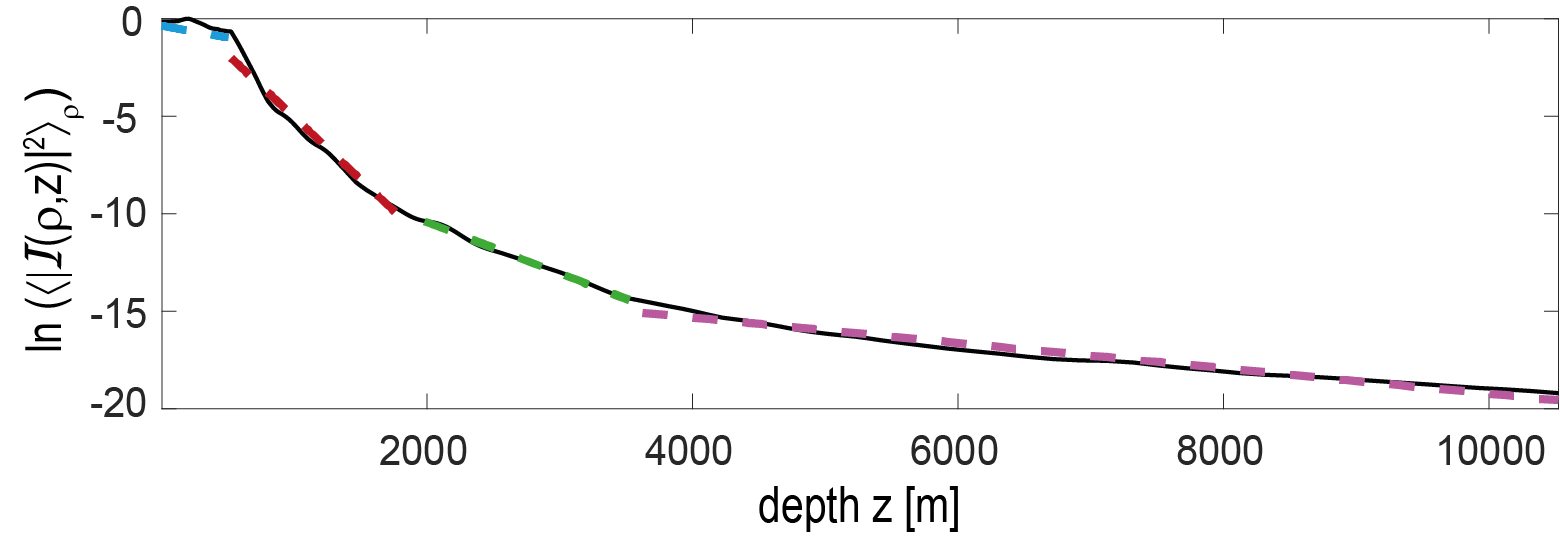}}
\caption{\textbf{Scattering intensity vs. depth}. The logarithm of the mean confocal intensity, $\langle |I'(\bm{\rho},z)^2\rangle_{\bm{\rho}} $, is plotted as a function of depth. The depth decay of intensity is fitted with an exponential curve over four depth ranges summarized in Tab.~1 of the accompanying paper.}
\label{normZ}
\end{figure*}

Supplementary Figure~\ref{normZ} shows the depth evolution of the raw confocal intensity. 

\section{Transmit point-spread function}

The PSF $H(\bm{\rho},\bm{\rho}_{\textrm{in/out}},z)$ can be expressed using the \textit{true} Green's function $G(\bm{\rho},\mathbf{g}_i,z)$ between the geophones and the focused basis ~\cite{lambert_reflection_2020}:
\begin{equation}
    H(\bm{\rho},\bm{\rho}_{\textrm{in/out}},z)=\sum_{\mathbf{g}_{i}} G(\bm{\rho},\mathbf{g}_i,z,f_0)G_0^{*}(\bm{\rho}_{\textrm{in/out}},\mathbf{g}_i,z,f_0) 
\end{equation}
This discrete equation can be rewritten under a continuous form as a function of a coordinate $\mathbf{u}$ describing the Earth surface at a depth origin $z=0$ defined by the average elevation of the seismic stations $\mathbf{g}_i$:
\begin{equation}
\label{PSF}
    H(\bm{\rho},\bm{\rho}_{\textrm{in/out}},z)=\int d\mathbf{u} \overline{O}(\mathbf{u})G(\bm{\rho},\mathbf{u},z,f_0)G_0^{*}(\bm{\rho}_{\textrm{in/out}},\mathbf{u},z,f_0) 
\end{equation}
with $\overline{O}(\mathbf{u})=N_g^{-1}\sum_i \delta (\mathbf{u}-\mathbf{g}_{||,i})$, the distribution of geophones, $N_g$, their number and $\delta$ the Dirac distribution. 

 In the absence of aberrations, \textit{i.e} if the wave velocity model is valid ($G\equiv G_0$), the expression of the PSF becomes under the Fresnel approximation:
 \begin{equation}
 \label{fresnel}
    H_0(\bm{\rho},\bm{\rho}_{\textrm{in/out}},z)= \rev{\frac{1}{z^2}} \exp \left [\rev{-i} \frac{k_0}{2z} \left (|\bm{\rho}|^2-|\bm{\rho}_{\textrm{in/out}}|^2 \right)\right] O \left ( \frac{\bm{\rho}-\bm{\rho}_{\textrm{in/out}}}{\lambda z}\right ) 
\end{equation}
In absence of aberration, the reference PSF $H_0$ is thus the product between: (\textit{i}) a geometrical spreading term; (\textit{ii}) a parabolic phase law that accounts for the curvature of the focused wave-front; (\textit{iii}) the Fourier transform $O$ of the geophone network aperture $\overline{O}$, such that $O \left ( \frac{\bm{\rho}-\bm{\rho}_{\textrm{in/out}}}{\lambda z}\right )=\int d\mathbf{u} \overline{O}(\mathbf{u})  \exp \left [i \frac{k_0}{z} \mathbf{u}.(\bm{\rho}-\bm{\rho}_{\textrm{in/out}})\right] $. The transverse dimension of the focal spot, $\delta \rho_u$, is then only limited by diffraction:
 \begin{equation}\label{deltarho_0}  
     \delta \rho_u (z) \sim \lambda_z / [2\sin[\theta_u(z)]
 \end{equation}
with $\theta_u$, the mean angle under which the geophone network is seen by the focusing point $\mathbf{r}_{\textrm{in/out}}$.
  
In the presence of aberrations, i.e., if the velocity model is
inaccurate, there is a mismatch between the true Green's matrix $\mathbf{G}$ and its model $\mathbf{G}_0$. If aberrations are moderate ($\delta \rho < \sqrt{\lambda z}$), they can be accounted for, at each depth $z$, by a phase screen $\overline{A}(\mathbf{u},z)$ at the Earth surface, such that
\begin{equation}
    G(\bm{\rho},\mathbf{u},z,f_0)=\overline{A}(\mathbf{u},z)G_0(\bm{\rho},\mathbf{u},z,f_0)
\end{equation}
Equation~\ref{PSF} then simplifies into:
\begin{equation}
\label{PSF2}
    H(\bm{\rho},\bm{\rho}_{\textrm{in/out}},z)=\int d\mathbf{u} \overline{F}(\mathbf{u},z)G_0 (\bm{\rho},\mathbf{u},z,f_0)G_0^{*}(\bm{\rho}_{\textrm{in/out}},\mathbf{u},z,f_0) 
\end{equation}
with $\overline{F}(\mathbf{u},z)=\overline{O}(\mathbf{u})\overline{A}(\mathbf{u},z)$, the overall transmittance that combines the array aperture $\overline{O}$ and the aberration phase screen $\overline{A}$. 
Under the Fresnel approximation,the previous equation becomes:
 \begin{equation*}
    H(\bm{\rho},\bm{\rho}_{\textrm{in/out}},z)=  \rev{\frac{1}{z^2}} \exp \left [\rev{-}i \frac{k_0}{2z} \left (|\bm{\rho}|^2-|\bm{\rho}_{\textrm{in/out}}|^2 \right)\right] \underbrace{\int d\mathbf{u} \overline{F}(\mathbf{u},z)  \exp \left [i \frac{k_0}{z} \mathbf{u}.(\bm{\rho}-\bm{\rho}_{\textrm{in/out}})\right] }_{F   \left (\frac{\bm{\rho}-\bm{\rho}_{\textrm{in/out}}}{\lambda z}\right )}
\end{equation*}
The PSF $H$ is thus the product of a parabolic phase law that results from the curvature of focused wave-fronts and a focusing function $F$, that results from the convolution between the network PSF $O$ that accounts for diffraction and the aberration PSF $A$ defined as the Fourier transform of the aberration transmittance $\overline{A}$:
 \begin{equation*}
    F   \left (\frac{\bm{\rho} -\bm{\rho}_{\textrm{in/out}}}{\lambda z}\right ) =O \otimes A \left (\frac{\bm{\rho} - \bm{\rho}_{\textrm{in/out}}}{\lambda z} , z \right )
\end{equation*}
where the symbol $\otimes $ stands for the convolution product.

\section{Reflection point spread function}

For a \textit{sparse} scattering medium like the present volcano in the considered frequency range, the reflectivity of the medium can be expressed as follows at each depth $z$:
\begin{equation}
\label{sparse}
\gamma(\bm{\rho},z)=\sum_s \gamma_s \delta(\bm{\rho}-\bm{\rho}_s(z)).
\end{equation}
with $\delta$, the Dirac distribution, $\bm{\rho}_s$, the position of the $s^{{th}}$ scatterers at depth $z$ and $\gamma_s$, its reflectivity. Injection Eq.~\ref{sparse} into Eq.~12 of the accompanying paper, the RPSF at each scatterer position $\bm{\rho}_s$ can be expressed as follows:
\begin{equation*}
    RSPF(\Delta \bm{\rho},\bm{\rho}_s,z)=\exp \left (i \frac{k_0}{z} |\Delta \bm{\rho}|^2  \right) F \left (\frac{\Delta \bm{\rho}}{\lambda z}\right )  F \left (-\frac{\Delta \bm{\rho}}{\lambda z} \right) .
\end{equation*}
Therefore, the energy spreading in the vicinity of each scatterer position shall enable one to probe the spatial extension of the PSF. As the scatterer positions are \textit{a priori} unknown, the RPSF is, in practice, probed by considering the antidiagonal whose common mid-point exhibits the maximum confocal signal.

\section{Iterative phase reversal from the Earth surface basis}

As highlighted in the previous section, the focused basis is the proper framework for imaging and quantification of focusing quality. However, a dual basis is a better framework to analyse and compensate for aberrations. In the accompanying paper, aberrations are unscrambled by: (\textit{i}) projecting the reflection matrix between the Earth surface and focused basis (Eq.~14 of the accompanying paper); (\textit{ii}) isolate wave distortions by comparing the reflected wave-field at the Earth surface with its ideal counterpart (Eq.~15). Injecting Eqs.~6 and 7 into Eqs.~14 and 15 leads to the following expression of the $\mathbf{D}-$matrix coefficients under the Fresnel approximation:
   \begin{multline}\label{Duout2}
     D(\mathbf{u}\out,\bm{\rho}\inp,z)= \overbrace{\overline{F}(\mathbf{u}\out,z) }^{\mbox{transmittance}} \\ \times \int d\bm{\rho} \underbrace{{\gamma}(\bm{\rho},z) \exp \left (\rev -i\frac{k_0}{z} \rev{ \left( \left |\bm{\rho} \right |^2 - \left |\bm{\rho}_{\textrm{in}} \right |^2\right)} \right) }_{\mbox{apparent reflectivity}} \underbrace{F \left (\frac{\bm{\rho}-\bm{\rho}_{\textrm{in}}}{\lambda z}\right )}_{\mbox{transmit PSF}} \underbrace{\exp \left (i\frac{k_0}{z}\mathbf{u}\out.(\bm{\rho}-\bm{\rho}_{\textrm{in}}) \right)}_{\mbox{angular de-scan}}
 \end{multline}

In a previous work~\cite{touma_distortion_2021}, the aberration transmittance was extracted through a singular value decomposition (SVD) of $\mathbf{D}_{\mathbf{u}\bm{\rho}}$ or, equivalently, an eigenvalue decomposition (EVD) of the correlation matrix $\mathbf{C}_{\mathbf{u}\mathbf{u}}=\mathbf{D}_{\mathbf{u}\bm{\rho}}\mathbf{D}_{\mathbf{u}\bm{\rho}}^{\dag}$~\cite{touma_distortion_2021,Lambert2022}. This result can be understood if we assume a point-like transmit PSF in the previous equation. In that case, we have $$  D(\mathbf{u}\out,\bm{\rho}\inp,z)= \overline{F}(\mathbf{u}\out)  {\gamma}(\bm{\rho}\inp,z).$$ The $\mathbf{D}$-matrix is then of rank 1 and its first singular vector $\mathbf{U}_1$ directly provides the aberration transmittance at depth $z$. 

Physically, the first eigenvector $\mathbf{U}_1$ is the result of a virtual iterative time reversal (ITR) experiment on a guide star whose reflectivity corresponds to the transmit PSF intensity~\cite{Lambert2022}. Mathematically, the time reversal invariant can be found by solving the following iterative relation~\cite{prada_eigenmodes_1994}:
\begin{equation}
    \lambda^{(n+1)}_1 \mathbf{U}_1^{(n+1)}=\mathbf{C}_{\mathbf{u}\mathbf{u}} \times \mathbf{U}_1^{(n)}
\end{equation} 
with $\mathbf{U}_1=\lim\limits_{n\rightarrow \infty}\mathbf{U}_1^{(n)}$ and $\lambda_1=\lim\limits_{n\rightarrow \infty}\lambda_1^{(n)}$, the first eigenvalue of $\mathbf{C}_{\mathbf{u}\mathbf{u}}$. ITR converges towards the wave-front that maximizes the energy backscattered by the virtual guide star. If this guide star is point like, $\mathbf{U}_1^{(n)}$ thus converges towards the aberration transmittance. However, in reality, the transmit PSF is of course not point-like and its blurring biases the estimation of the aberration transmittance with a time reversal invariant that concentrates on the central part of the geophone array and vanishes on its edge~\cite{Lambert2022}. 

To circumvent this problem, a related approach consists in an iterative phase reversal (IPR) process~\cite{Bureau2023,Najar2023} that forces a uniform amplitude for the phase reversal invariant $\mathbf{W}$ (Methods). Mathematically, the success of IPR can be explained by expressing the correlation matrix $\mathbf{C}_{\mathbf{u}\mathbf{u}}$. In the speckle regime~\cite{lambert_ultrasound_2021} or in a sparse medium made of a few point-like reflectors, its coefficients can be expressed as follows:
\begin{equation}
\label{Cu}
C(\mathbf{u}\out,\mathbf{u}'\out,z)= \overline{F}(\mathbf{u}\out,z) \overline{F}^*(\mathbf{u}'\out,z) \left [ \overline{F}  \stackrel{\mathbf{u}\out}{\otimes} \overline{F}\right ] (\mathbf{u}\out-\mathbf{u}'\out,z)
\end{equation}
The correlation term of the right hand side results from the Fourier transform of the input focal spot intensity distribution $|F\left ( {\bm{\rho}-\bm{\rho}\inp}/\lambda z\right )|^2$. This formulation is reminiscent of the Van Cittert Zernike theorem for an aberrating layer, which links the spatial correlation of a wavefield to the Fourier transform of the intensity distribution from the virtual guide stars (here the input focal spots). In other words, the support of the coherence function scales as the inverse of the input focal spot size. Injecting Eq.~\ref{Cu} into Eq.~21 leads to the following equation:
\begin{equation}
    W(\mathbf{u}\out,z)=\exp \left ( j \mbox{arg} \left \lbrace \overline{F}(\mathbf{u}\out,z) \sum_{\mathbf{u}'\out} W(\mathbf{u}'\out,z) \overline{F}^*(\mathbf{u}'\out,z) \left [ \overline{F}  \stackrel{\mathbf{u}\out}{\otimes} \overline{F}\right ] (\mathbf{u}\out-\mathbf{u}'\out , z) \right \rbrace \right ).
\end{equation}
For a real autocorrelation function $ \overline{F}  {\otimes} \overline{F}$, or equivalently, a symmetric input PSF $|F|^2$, the solution of the previous equation is
\begin{equation}
    W(\mathbf{u}\out, z)=\overline{F}(\mathbf{u}\out,z).
\end{equation}
If the previous condition is not fulfilled, the estimation of the aberration transmittance suffers from a bias that can be reduced by iterating the aberration correction process, thereby gradually reducing the size of the virtual guide star and flattening the autocorrelation function $\overline{F}  {\otimes} \overline{F}$.

\section{Iterative phase reversal driven from the $\mathbf{k}$-space}

To beat diffraction, we exploit the parabolic phase law exhibited by the transmit PSF (Eq.~7)  highlighted by the phase of the RPSF in Fig.~4h of the accompanying paper. To that aim, the $\mathbf{R}-$matrix is investigated from the $\mathbf{k}-$space.

As previously done for the geophone basis, the first step consists in a projection of the focused reflection matrix in the plane wave basis (Eq.~17).
Then, the $\mathbf{D}-$matrix is built by isolating the mismatch between each reflected wave-field and a reference wave-field that would be obtained for a point-like guide star at $\bm{r}\inp$ (Eq.~19). To derive an expression for the $\mathbf{D}$-matrix coefficients in the $\mathbf{k}-$space, one can inject Eqs.~6, \ref{fresnel}, 17 and 18 into Eq.~19. Assuming that aberrations have been fully compensated in the geophone basis ($F \equiv O$ and $\overline{F} \equiv \overline{O}$), this expression writes as follows:
 \begin{multline}\label{Dkout2}
     {D}(\mathbf{k}\out,\bm{\rho}\inp,z)=   \int d \mathbf{\bm{\rho}} \overbrace{K(\mathbf{k}\out,\bm{\rho},z) }^{\mbox{transmittance}}   
   \overbrace{ \gamma(\bm{\rho},z) \exp \left (- i\frac{k_0}{2z} |\bm{\rho}|^2 \right ) }^{\mbox{reflectivity}}\\ \underbrace{O\left ( \frac{\bm{\rho} - \bm{\rho}\inp}{\lambda z}\right )  \exp \left (i\frac{k_0}{2z} |\bm{\rho}\inp|^2 \right ) }_{\mbox{input PSF}} \underbrace{\exp \left [{\rev{-}i\mathbf{k}\out.(\bm{\rho}- \bm{\rho}\inp)}\right] }_{\mbox{lateral de-scan}},
 \end{multline}
 with
 \begin{equation}
\label{K}
   K(\mathbf{k}\out,\bm{\rho},z)= \mathbb{1}_{|\mathbf{k}\out|<k_0} \left \lbrace \overline{O}(\bm{\rho} + \lambda z \mathbf{k}\out ) \stackrel{\mathbf{k}\out} {\otimes}  \exp \left (i \pi \lambda z |\mathbf{k}\out|^2   \right) \right \rbrace,
 \end{equation}
the aperture transmittance projected in the $\mathbf{k}$-space. The symbol $\mathbb{1}_{|\mathbf{k}\out|<k_0}$ accounts for the low pass filter operated by diffraction in the spatial frequency domain: Only spatial frequencies $\mathbf{k}\out$ whose magnitudes are smaller than $|k_0|$ can propagate into the Earth; higher spatial frequency components are evancescent and cannot propagate into the Earth beyond a wavelength.

If we compare Eq.~\ref{Dkout2} with the $\mathbf{D}-$matrix expression in the geophone basis (Eq.~\ref{Duout2}), the main difference lies in the fact that the aperture transmittance is no longer isoplanatic from the $\mathbf{k}-$space. Indeed, the angular component of the wave-field recorded by the geophone network and induced by one scatterer in the field-of-view strongly depends on its position, as shown by the term $\overline{O}(\bm{\rho} + \lambda z \mathbf{k}\out )$ in Eq.~\ref{K} and highlighted by Fig.~5a of the accompanying paper.  While this would be an issue in the speckle regime (random reflectivity), this property can become an asset in a sparse scattering medium (Eq.~\ref{sparse}). Under this assumption, Eq.~\ref{Dkout2} becomes:
 \begin{multline}\label{Dkout3}
     \mathbf{D}(\mathbf{k}\out,\bm{\rho}\inp,z)=   \sum_{s}  K(\mathbf{k}\out,\bm{\rho}_s,z)  
   { \gamma_s \exp \left (- i\frac{k_0}{2z} |\bm{\rho}_s(z)|^2 \right ) }  O\left ( \frac{\bm{\rho}_s(z) - \bm{\rho}\inp}{\lambda z}\right )\\  \exp \left (i\frac{k_0}{2z} |\bm{\rho}\inp|^2 \right )  \exp \left [{\rev{-}  i\mathbf{k}\out.(\bm{\rho}_s(z)- \bm{\rho}\inp)}\right].
 \end{multline}
The coefficients of the associated correlation matrix, $\mathbf{C}_{\mathbf{k}\mathbf{k}}(z)=\mathbf{D}_{\mathbf{k}\bm{\rho}}(z)\mathbf{D}^{\dag}_{\mathbf{k}\bm{\rho}}(z)$, are given by:
\begin{equation}
\label{Ck}
C(\mathbf{k}\out,\mathbf{k}'\out,z)= \sum_{s} |\gamma_s|^2 K(\mathbf{k}\out,\bm{\rho}_s,z) K^*(\mathbf{k'}\out,\bm{\rho}_s,z) \left [ \overline{O}  \stackrel{\mathbf{k}\out}{\otimes} \overline{O}\right ] (\mathbf{k}\out-\mathbf{k}'\out)
\end{equation}
For analytical tractability, we will consider, in first approximation, the correlation term $ \overline{O}  \stackrel{\mathbf{k}\out}{\otimes} \overline{O}$ as constant:
\begin{equation}
\label{Ck2}
C(\mathbf{k}\out,\mathbf{k}'\out,z) \simeq \sum_{s} |\gamma_s|^2 K(\mathbf{k}\out,\bm{\rho}_s) K^*(\mathbf{k'}\out,\bm{\rho}_s)
\end{equation}
Under this assumption and provided that the scatterers belong to different resolution cells ($\mathbf{K}^{\dag}\mathbf{K}\propto  \mathbf{I}$, with $\mathbf{I}$ the identity matrix), Eq.~\ref{Ck2} has the form of an eigenvalue decomposition of $\mathbf{C}_{\mathbf{k}\mathbf{k}}$:
\begin{equation}
\label{evd}
    C(\mathbf{k}\out,\mathbf{k}'\out)=\sum_s \sigma_s^2 U_s(\mathbf{k}\out)U_s^*(\mathbf{k}'\out)
\end{equation}
with $\sigma_s$, the singular values of $\mathbf{D}_{\mathbf{k},\bm{\rho}}$, and $\mathbf{U}_s=[U_s(\mathbf{k}\out)]$, the eigenvectors of $\mathbf{C}_{\mathbf{k}\mathbf{k}}$, or equivalently the output singular vectors of $\mathbf{D}_{\mathbf{k}\bm{\rho}}(z)$. The identification of Eqs.~\ref{Ck2} and \ref{evd} shows a one-to-one association between each eigenstate of $\mathbf{C}_{\mathbf{k}\mathbf{k}}$ and each scatterer. The singular values $\sigma_s$ are proportional to each scatterer reflectivity $\gamma_s$, while each eigenvector $\mathbf{U}_s$ provides the far-field transmittance $\mathbf{K}(\mathbf{k}\out,\bm{\rho}_s)$ of each scatterer. 

\clearpage 

\begin{figure*}    
\centerline{\includegraphics[width = \textwidth]{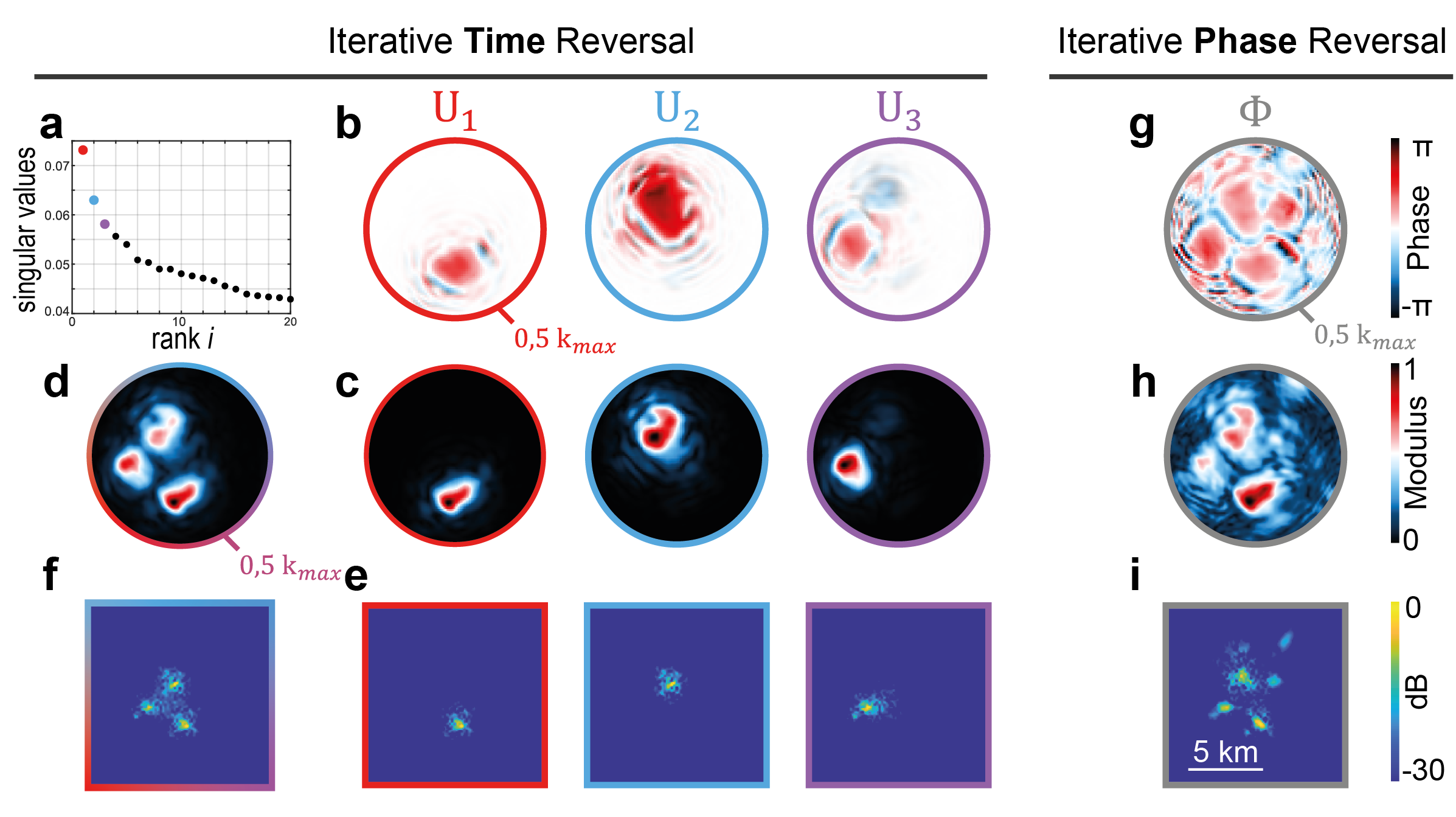}}
\caption{\textbf{Comparison between iterative time reversal and iterative phase reversal processes for aberration phase law extraction} - Illustration at depth $z = 6.9$ km. \textbf{a}, Singular value histogram of the distortion matrix $\mathbf{D_{kr}}$. \textbf{b}-\textbf{c}, Phase (\textbf{b}) and modulus (\textbf{c})  of the first three singular vectors $\mathbf{U}_s = [U_s(\mathbf{k}_{\textrm{out}})]$. \textbf{d}, Sum of the modulus of the first three eigenvectors. \textbf{e}, Images obtained using the phase of the first three singular vectors $\mathbf{U}_s$ as the focusing law. \textbf{f}, Sum of the images displayed in (\textbf{e}). \textbf{g}-\textbf{h}, Phase $\phi$ of the IPR invariant $\mathbf{W}$ (\textbf{g}) and associated angular spectrum obtained by considering the modulus of $\mathbf{C_{kk}} \times \mathbf{W}$ (\textbf{h}). \textbf{i}, Confocal image using the phase conjugate of $\mathbf{W}$ as the focusing law. }
\label{ITR_IPR}
\end{figure*}

Supplementary Fig.~\ref{ITR_IPR} confirms this conjecture by showing the result of the ITR processing applied to $   \mathbf{D}_{\mathbf{k}\bm{\rho}}(z)$ at depth $z=6.9$ km. As seen before, this process is mathematically equivalent to the SVD of  $ \mathbf{D}_{\mathbf{k}\bm{\rho}}(z)$. Supplementary Fig.~\ref{ITR_IPR}a shows the singular value spectrum of $\mathbf{D}_{\mathbf{k}\bm{\rho}}(z)$ dominated by three singular values eigenvalues. Each corresponding eigenvector $\mathbf{U}_s$ covers a distinct part of the $\mathbf{k}$-space (Supplementary Fig.~\ref{ITR_IPR}d), as predicted by the term $\overline{O}(\bm{\rho}_s + \lambda z \mathbf{k}\out )$ in Eq.~\ref{K}. The phase of these eigenvectors (Supplementary Fig.~\ref{ITR_IPR}b) shows the Fresnel rings corresponding to the parabolic phase term $\exp \left (i \pi \lambda z |\mathbf{k}\out|^2\right )$ in Eq.~\ref{K}. 

Note that this observation also enables to revisit the results of Touma et al.~\cite{touma_distortion_2021} that showed a similar feature in the fault area of San Jacinto. As shown in that previous paper, the phase conjugate of each eigenstate can provide the focusing law to image each scatterer (Supplementary Fig.~\ref{ITR_IPR}f). A compound image can be built by combining the result provided by each eigenstate. However, this approach only provides an image of the main structures of the volcano. As we will see further, it actually fails in highlighting smaller reflectors.

The IPR process that we previously introduced above in the geophone basis can provide a much more complete view of the inner volcano structure. By forcing a transmittance estimation with the same weight over the whole $\mathbf{k}-$space, the resulting focusing law $\mathbf{W}$ (Supplementary Fig.~\ref{ITR_IPR}g) can address simultaneously all scatterers in the field-of-view. The angular spectrum addressed by this focusing law can be estimated by considering the modulus of the vector $\mathbf{C_{kk}}\times \mathbf{W}$ (Supplementary Fig.~\ref{ITR_IPR}h). The comparison with the angular spectrum covered by the three first eigenvectors of $\mathbf{C}_{\mathbf{k},\mathbf{k}}(z)$ (Supplementary Fig.~\ref{ITR_IPR}d) shows the benefit of IPR for tailoring a focusing law operating over the whole $\mathbf{k}$-space. Also applied at input of the $\mathbf{R}-$matrix, the focusing law derived by IPR leads to the confocal image displayed in Supplementary Fig.~\ref{ITR_IPR}i. Compared with the ITR process that only focuses on the three main reflectors at the considered depth {(Supplementary Fig.~\ref{ITR_IPR}e)}, the IPR algorithm provides a full-field image of the subsoil highlighting six main structures at the same depth {(Supplementary Fig.~\ref{ITR_IPR}i)}.

\begin{figure*}    
\centerline{\includegraphics[width = 10 cm]{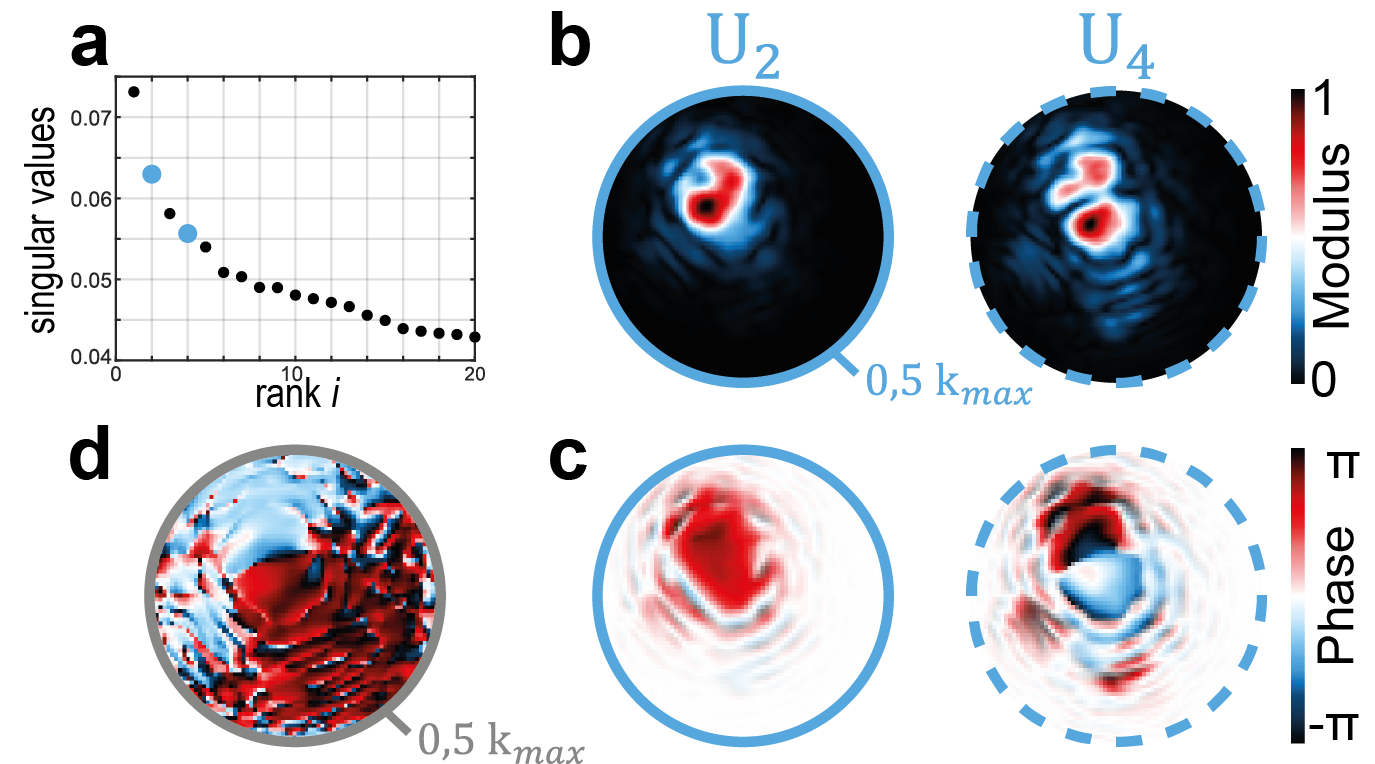}}
\caption{\textbf{Limit of iterative time reversal processing} - Illustration at depth $z = 6.9$ km. \textbf{a}, Singular value histogram of the distortion matrix $\mathbf{D_{kr}}$. \textbf{b}-\textbf{c}, Modulus (\textbf{b}) and phase (\textbf{c}) of the second and fourth singular vectors, $\mathbf{U}_2$ and $\mathbf{U}_4$. \textbf{d}, Phase of the Hadamard product between $\mathbf{U}_2$ and $\mathbf{U}_4^*$. }
\label{ITR2}
\end{figure*}
One could argue that this difference comes from the fact that we did not consider enough eigenstates in the ITR process. However, the higher-order eigenstates cannot be used for imaging purposes~\cite{touma_distortion_2021}. Supplementary Fig.~\ref{ITR2} illustrates this fact by showing the fourth eigenstate $\mathbf{U}_4$ of $\mathbf{D}_{\mathbf{k}\bm{\rho}}$. Its support emerges in the same angular range as $\mathbf{U}_2$. The link between the second and fourth eigenstates is confirmed by Supplementary Fig.~\ref{ITR2}d that displays the phase difference between $\mathbf{U}_{4}$ and $\mathbf{U}_2$. $\mathbf{U}_{4}$ is an higher-order eigenstate associated with the same reflector as $\mathbf{U}_{2}$. Each reflector gives actually rise to a set of eigenmodes induced by the autocorrelation term in Eq.~\ref{Ck}. Only the fundamental modes corresponding to the highest singular values can be considered for imaging. Higher-order eigenmodes as $\mathbf{U}_{4}$ in Supplementary Fig.~\ref{ITR2} correspond to smaller singular values and give rise to distorted PSFs~\cite{Lambert2022}. They cannot be used for imaging but they pollute the singular value spectrum of $\mathbf{D}_{\mathbf{k}\bm{\rho}}$. Hence they can prevent from imaging scatterers of smaller reflectivity within the framework of an ITR process. The proposed IPR process allows to circumvent this limit by finding a phase reversal invariant over the whole angular spectrum (Supplementary Fig.~\ref{ITR_IPR}g).

\section{Spatial resolution of the final image}

As highlighted by the final confocal image (Figs.~5d and e of the accompanying paper) and corresponding RPSF (Fig.~5f), the IPR algorithm driven from the $\mathbf{k}-$space leads to a resolution of the order of $\lambda/2$ much thinner than the usual diffraction limit dictated by the geophone aperture: $\delta \rho_u \sim \lambda z/d_0$. Mathematically, this can be understood by the convolution product between the geophone aperture and Fresnel rings exhibited by the transmittance $K(\mathbf{k}\out,\rho_s)$ in the  $\mathbf{k}$-space (Eq.~\ref{K}). These Fresnel rings originate from the parabolic phase law exhibited by the focal spots in real space (Eq.~13). Encoded in the secondary lobes of the PSF, this Fresnel phase law exhibits spatial frequency components from $0$ to $k_0$. When properly realigned in phase, those frequency components lead to a corrected PSF whose extension spans over $\lambda/2$ instead of the usual aperture limited resolution $\Delta \rho_u \sim \lambda z/d_0$.

\section{Depth evolution of the maximum confocal signal}
\label{sec10}

\begin{figure*}    
\centerline{\includegraphics[width = 14 cm]{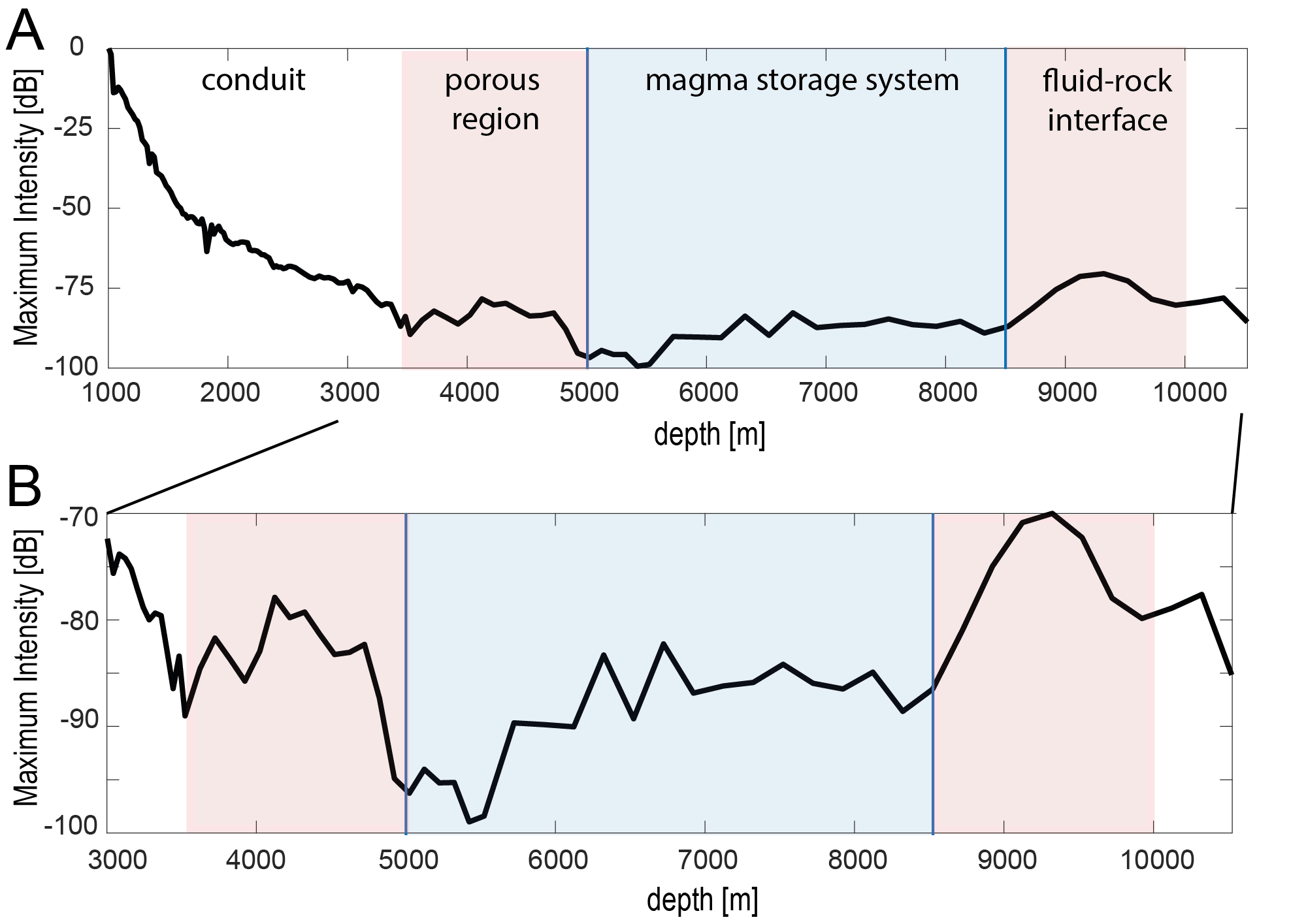}}
\caption{\textbf{Depth evolution of the maximum confocal intensity at each depth}: \textbf{a}, between $z=1$ and 10.5 km; \textbf{b}, between $z=3$ and 10.5 km. }
\label{depth}
\end{figure*}

Supplementary Fig.~\ref{depth} shows the depth evolution of the maximum confocal signal at each depth without normalization. This curve is plotted at the end of the matrix imaging process. The compensation of aberrations and diffraction operated by matrix imaging allows a finer analysis than the preliminary fit of the confocal signal provided in Supplementary Fig.~1. It shows different behaviors in each main part of the volcano. While a strong attenuation is observed in the upper part of the volcano ($z=$0-3.5 km, see also Supplementary Fig.~\ref{normZ}), the deeper part of the volcano (Supplementary Fig.~\ref{depth}b) exhibits fluctuations around a relatively constant reflectivity. The magma storage zone ($z$=5-8.5 km) shows a weaker reflectivity probably due the presence of extended magma volumes. This region is surrounded by two areas of larger reflectivity around 4 and 9 km. This larger confocal signal is probably a manifestation of the important impedance mismatch existing at the boundaries of the magma storage zone (fluid-rock interface). \alex{Above the outer carapace of the magma storage system, there are lenses of supercritical acid fluids/brines, and then closer to the surface these supercritical fluids become zones with gases and/or liquid hydrothermal fluids that are present in the pores of the host-rocks, along special zones of porosity-permeability~\cite{Moretti2020}. On the one hand, this porous region may account for the enhancement of the confocal signal observed between $z=3.5$ and 5 km in Supplementary Fig.~\ref{depth}. On the other hand, the increase of reflectivity at the bottom of the magma storage system ($z=8.5-10$ km) is a priori due to the back-reflection echo induced by the interface between eruptive melt of the magma storage system and the deeper host-rock.}   

\end{document}